\documentclass[10pt]{article}
\usepackage{graphicx}
\usepackage{rotating}
\usepackage{booktabs}
\renewcommand{\baselinestretch}{1.}
\begin{document}

\begin{center}
{\Large \bf Excitation functions of parameters in Erlang
distribution, Schwinger mechanism, and Tsallis statistics in RHIC
BES program}

\vskip1.0cm

Li-Na Gao$^{a}$, Fu-Hu Liu$^{a,}${\footnote{E-mail:
fuhuliu@163.com; fuhuliu@sxu.edu.cn}}, and Roy A.
Lacey$^{b,}${\footnote{E-mail: Roy.Lacey@Stonybrook.edu}}

{\small\it $^a$Institute of Theoretical Physics, Shanxi
University, Taiyuan, Shanxi 030006, China

$^b$Departments of Chemistry \& Physics, Stony Brook University,
Stony Brook, NY 11794, USA}
\end{center}

\vskip1.0cm

{\bf Abstract:} Experimental results of the transverse momentum
distributions of $\phi$ mesons and $\Omega$ hyperons produced in
gold-gold (Au-Au) collisions with different centrality intervals,
measured by the STAR Collaboration at different energies (7.7,
11.5, 19.6, 27, and 39 GeV) in the beam energy scan (BES) program
at the relativistic heavy ion collider (RHIC), are approximately
described by the single Erlang distribution and the two-component
Schwinger mechanism. Meanwhile, the STAR experimental transverse
momentum distributions of negatively charged particles, produced
in Au-Au collisions at RHIC BES energies, are approximately
described by the two-component Erlang distribution and the single
Tsallis statistics. The excitation functions of free parameters
are obtained from the fit to the experimental data. A weak softest
point in the string tension in $\Omega$ hyperon spectra is
observed at 7.7 GeV.
\\

\\


\vskip1.0cm

{\section{Introduction}}

High energy nucleus-nucleus (heavy ion) collisions at the
relativistic heavy ion collider (RHIC) [1--3] and the large hadron
collider (LHC) [4, 5] can provide the environment and condition of
high temperature and density, where a new state of matter, namely
the quark-gluon plasma (QGP), is expected to form [6--8]. Even at
the top center-of-mass energy (the maximum $\sqrt{s_{NN}}=19.4$
GeV) of the super proton synchrotron (SPS), the equation of state
(EoS) is possibly different from that at lower energy [9]. Both
the RHIC and SPS have performed the beam energy scan (BES)
programs, which give us opportunities to search for the onset of
deconfinement of quarks and gluons, the critical point of phase
transition from QGP to hadronic matter, and the softest points in
the EoS in nucleus-nucleus collisions. The onset of quark
deconfinement should be possibly the critical point of QGP phase
transition. Although there is a relation between the softest
points in the EoS and the critical point of QGP phase transition,
they are not sure the same. In addition, different quantities
result in different softest points, and the critical point should
be exclusive. To reveal the relation between the softest points
and critical point is still an open question.

Generally, the RHIC and its BES energies ($\sqrt{s_{NN}}=7.7$--200
GeV) [10--12] are larger than the SPS and its BES energies
($\sqrt{s_{NN}}=6.6$--19.4 GeV) [13--15], though they have an
overlap each other. The STAR Collaboration has been performing the
RHIC BES program [10--12], and the NA61/SHINE Collaboration has
been performing the SPS BES program [13--15]. Some interesting
results involved to the softest points in some excitation
functions in the EoS have been reported. For example, the works
which study excitation functions of ratio of positive kaon number
to pions ($K^+/\pi^+$) [16--18], chemical freeze-out temperature
($T_{ch}$) [17, 18], mean transverse mass minus rest mass
($\langle m_T \rangle -m_0$) [17], and ratio of experimental
negative pion rapidity ($y$) distribution width to Landau
hydrodynamic model prediction [$\sigma_y(\pi^-)/\sigma_y({\rm
hydro})]$ [18] show the softest point being around
$\sqrt{s_{NN}}=7$--8 GeV. By using the Landau hydrodynamic model
and the ultra-relativistic quantum molecular dynamics hybrid
approach, an analysis based on rapidity distribution and squared
speed-of-sound ($c^2_s$) extraction shows the softest point
locating in the energy range from 4 to 9 GeV [19, 20].

Recently, a wiggle in the excitation function of a specific
reduced curvature, $C_y$, of the net-proton rapidity distribution
at midrapidity is expected in the energy range from 4 to 8 GeV
[21, 22], where $C_y=(y^3_{\rm beam}d^3N/dy^3)_{y=0}/(y_{\rm
beam}dN/dy)_{y=0}$, $y_{\rm beam}$ denotes the beam rapidity, and
$N$ denotes the number of considered particles. A local weak
wiggle in the $C_y$ excitation function is exhibited in the range
from 8 to 12 GeV. It is shown that there are more than one softest
points in the EoS from 4 to 12 GeV. Our recent work which based on
rapidity distribution and $c^2_s$ extraction shows the softest
point being around 8.8 GeV [23], and there is a jump (from $\leq
0.3$ to 1/3--1/2) in $c^2_s$ when the energy changes from 17.3 to
19.6 GeV. These phenomenons render that the interacting process is
complex in the considered energy range. However, the searching for
the onset of quark deconfinement and the critical point of QGP
phase transition is not simple [24]. More analyses are needed in
the study of the relation between the softest points and critical
point.

The present work does not intend to reveal the relation between
the softest points in the EoS and the critical points of QGP phase
transition. Instead, we hope to search for more information of the
softest points. In this paper, we study the transverse momentum
spectra of different particles ($\phi$ mesons, $\Omega$ hyperons,
and negatively charged particles) produced in gold-gold (Au-Au)
collisions at a few RHIC BES energies (7.7, 11.5, 19.6, 27, and 39
GeV) by using the single (or two-component) Erlang distribution
[25--27] for both hard and soft processes, two-component Schwinger
mechanism [28--31] for hard process only, and single Tsallis
statistics [32--34] for soft process only. The results calculated
by the two methods (Schwinger and Erlang, or Tsallis and Erlang)
are compared with the experimental data of the STAR Collaboration
[35, 36].

The structure of the present work is as follows. The model and
method are shortly described in section 2. Results and discussion
are given in section 3. In section 4, we summarize our main
observations and conclusions.
\\

{\section{The model and method}}

{\it Firstly}, we discuss uniformly hard and soft collision
processes in the framework of the multisource thermal model
[25--27]. According to the model, a given particle is produced in
the collision process where a few partons taken part in. Hard
process contains two or three partons which are valence quarks.
Soft process contains usually two or more partons which are gluons
and sea quarks. Each (the $i$-th) parton is assumed to contribute
to an exponential function [$f_i(p_t)$] of transverse momentum
($p_t$) distribution. Let $\langle p_t \rangle$ denotes the mean
transverse momentum contributed by each (the $i$-th) parton, we
have
\begin{equation}
f_i(p_t)=\frac{1}{\langle p_t \rangle} \exp \bigg(
-\frac{p_t}{\langle p_t \rangle} \bigg).
\end{equation}
The contribution of all $n$ partons which taken part in the
collision process is the folding of $n$ exponential functions
[25--27]. We have the transverse momentum ($p_T$) distribution
$f(p_T)$ of final-state particles to be the Erlang distribution
\begin{equation}
f(p_T)=\frac{p^{n-1}_T}{(n-1)!\langle p_t \rangle^n} \exp \bigg(
-\frac{p_T}{\langle p_t \rangle} \bigg),
\end{equation}
which has the mean transverse momentum $\langle p_T \rangle = n
\langle p_t \rangle$.

{\it Secondly}, the production of particles with heavy mass in
hard process is also described by the Schwinger mechanism
[28--31]. Let $\kappa$ denote the string tension between partons
which form a given final-state particle. Each (the $i$-th) parton
has and then contributes to the given particle a Gaussian function
of transverse momentum distribution
\begin{equation}
f_i(p_t)=\frac{1}{\sqrt{\kappa}} \exp \bigg( -\frac{\pi
p^2_t}{\kappa} \bigg).
\end{equation}
The transverse momentum distribution contributed by the string
(two partons) is the folding of two Gaussian functions. That is
\begin{equation}
f(p_T)=\int^{p_T}_0 f_1(p_t)f_2(p_T-p_t)dp_t = \frac{1}{\kappa}
\int^{p_T}_0 \exp \bigg\{ -\frac{\pi [p^2_t+(p_T-p_t)^2]}{\kappa}
\bigg\} dp_t.
\end{equation}

{\it Thirdly}, the Tsallis statistics which has more than one
function forms [32--34] is also used to describe the thermal
production of particles in soft process. We consider the simplest
form of the Tsallis transverse momentum distribution at
mid-rapidity
\begin{equation}
f(p_T)=C_T p_T \sqrt{p^2_T+m^2_0} \bigg[ 1+\frac{q-1}{T} \Big(
\sqrt{p^2_T+m^2_0}-\mu \Big) \bigg]^{-q/(q-1)},
\end{equation}
where $T$ is the temperature which describes averagely a few local
sources (equilibrium states), $q$ is the entropy index which
describes the degree of non-equilibrium among different local
sources, $\mu$ is the chemical potential which is related to
$\sqrt{s_{NN}}$ [37], $m_0$ is the rest mass of the considered
particle, and $C_T$ is the normalization constant which is related
to other parameters.

In the Monte Carlo method, let $r_i$ denote random numbers in
[0,1]. The Erlang distribution results in
\begin{equation}
p_T=-\langle p_t \rangle \sum_{i=1}^n \ln r_i = -\langle p_t
\rangle \ln \prod_{i=1}^n r_i.
\end{equation}
Although both $-\langle p_t \rangle \sum_{i=1}^n \ln r_i$ and
$-\langle p_t \rangle \ln \prod_{i=1}^n r_i$ can be used, we would
rather use $-\langle p_t \rangle \sum_{i=1}^n \ln r_i$ than
$-\langle p_t \rangle \ln \prod_{i=1}^n r_i$ due to $\prod_{i=1}^n
r_i$ being probably a too small value for the calculation when $n$
is large enough. Similarly, let $R_{1,2,3,4,5}$ denote random
numbers in [0,1]. The Schwinger mechanism results in
\begin{equation}
p_T=\sqrt{\frac{\kappa}{\pi}} \bigg[ \sqrt{-\ln R_1} \cos(2\pi
R_2) +\sqrt{-\ln R_3} \cos(2\pi R_4) \bigg].
\end{equation}
The Tsallis distribution satisfies
\begin{equation}
\int^{p_T}_0 f(p_T) dp_T <R_5< \int^{p_T+dp_T}_0 f(p_T) dp_T.
\end{equation}
Sometimes, the experimental $p_T$ distribution is the sum of two
components with different weights. This means that we need the
two-component function in some cases. Although we show the Monte
Carlo method of the calculations, the analytical method is used in
the present work, unless $(n-1)!$ is too large for the
calculations.

It should be noted that the Erlang distribution shows flexible
forms which are the foldings of two, three, $\cdots$, or multiple
exponential functions, the Schwinger mechanism shows an inflexible
form which is the folding of two Gaussian functions, while the
Tsallis statistics shows relative flexible form which is the sum
of two or three standard distributions [38]. The parameter
$\langle p_t \rangle$ in the Erlang distribution renders the mean
transverse momentum contributed by each parton, the parameter
$\kappa$ in the Schwinger mechanism renders the string tension
between two collision partons, while the parameter $T$ in the
Tsallis statistics renders the excitation degree of interacting
system. Because of the parton number in a string being fixed to
two, the Schwinger mechanism is not suitable in analysis of soft
process in which there are two or more partons, unless we consider
multiple strings and the folding of multiple Gaussian functions in
the process. In the case of using two or more components, the
parameter value can be obtained by the average weighted those of
different components. As a thermal description, the Tsallis
statistics is not suitable in analysis of hard process in which
heavy particles are produced by direct collisions between two or
among three partons.
\\

{\section{Results and discussion}}

Figure 1 presents the transverse momentum distributions,
$d^2N/(2\pi N_{evt}p_T dy dp_T)$, of $\phi$ mesons produced in
mid-rapidity ($|y|<0.5$) in Au-Au collisions at five RHIC BES
energies: (a) 7.7, (b) 11.5, (c) 19.6, (d) 27, and (e) 39 GeV,
where $N_{evt}$ denotes the event number and $\phi$ mesons are
assumed to produce in hard process. The symbols represent the
experimental data of the STAR Collaboration [35] in different
centrality intervals, which are scaled by different amounts marked
in the panel. The dashed and solid curves are our results
calculated by using the single Erlang distribution and the
two-component Schwinger mechanism respectively. The values of $n$
for all the cases are taken to be 3. Other related parameter
values are listed in Table 1 with values of $\chi^2$ per degree of
freedom ($\chi^2$/dof), where $k_{S1}$ denotes the relative
contribution of the first component in the two-component Schwinger
mechanism, and $\kappa_1$, $\kappa_2$, and $\langle p_t \rangle$,
are the main free parameters in the two functions. One can see
that both the single Erlang distribution and the two-component
Schwinger mechanism describe approximately the experimental
transverse momentum distributions of $\phi$ mesons produced in
mid-rapidity in Au-Au collisions at the RHIC BES in most cases.

We would like to point out that the experimental data in narrow
transverse momentum range should be described by the one-component
formula of the Schwinger mechanism. As an example, the dotted
curves in Figure 1(a) are the results of the single Schwinger
mechanism with $\kappa=1.462\pm0.283$, $1.435\pm0.275$,
$1.300\pm0.186$, $1.156\pm0.169$, $1.070\pm0.147$, and
$0.992\pm0.163$ GeV/fm, resulting in $\chi^2$/dof=0.084, 0.123,
0.200, 0.056, 0.044, and 0.421, respectively, when the centrality
interval changes from 0--10\% to 60--80\%. Although less
parameters are used in the fit of one-component formula, it does
not describe the data in wide transverse momentum range.

Figure 2 is similar to Figure 1, but it shows the results,
$d^2N/(2\pi N_{evt}p_T dy dp_T)$, of $\Omega^-$ (solid symbols)
and $\bar \Omega^+$ (open symbols divided by 2) in different
centrality intervals, where the symbols represent the data of the
STAR Collaboration [35] and $\Omega$ hyperons are assumed to
produce in hard process. At the same time, the result of the
one-component Schwinger mechanism is not available. The values of
$n$ for all the cases are taken to be 3. Other related parameters
are listed in Table 2 with values of $\chi^2$/dof. One can see
that both the single Erlang distribution and the two-component
Schwinger mechanism describe approximately the experimental data
of $\Omega^-$ and $\bar \Omega^+$ produced in mid-rapidity region
in Au-Au collisions at the RHIC BES in most cases.

\begin{figure}
\hskip-1.0cm \begin{center}
\includegraphics[width=11.0cm]{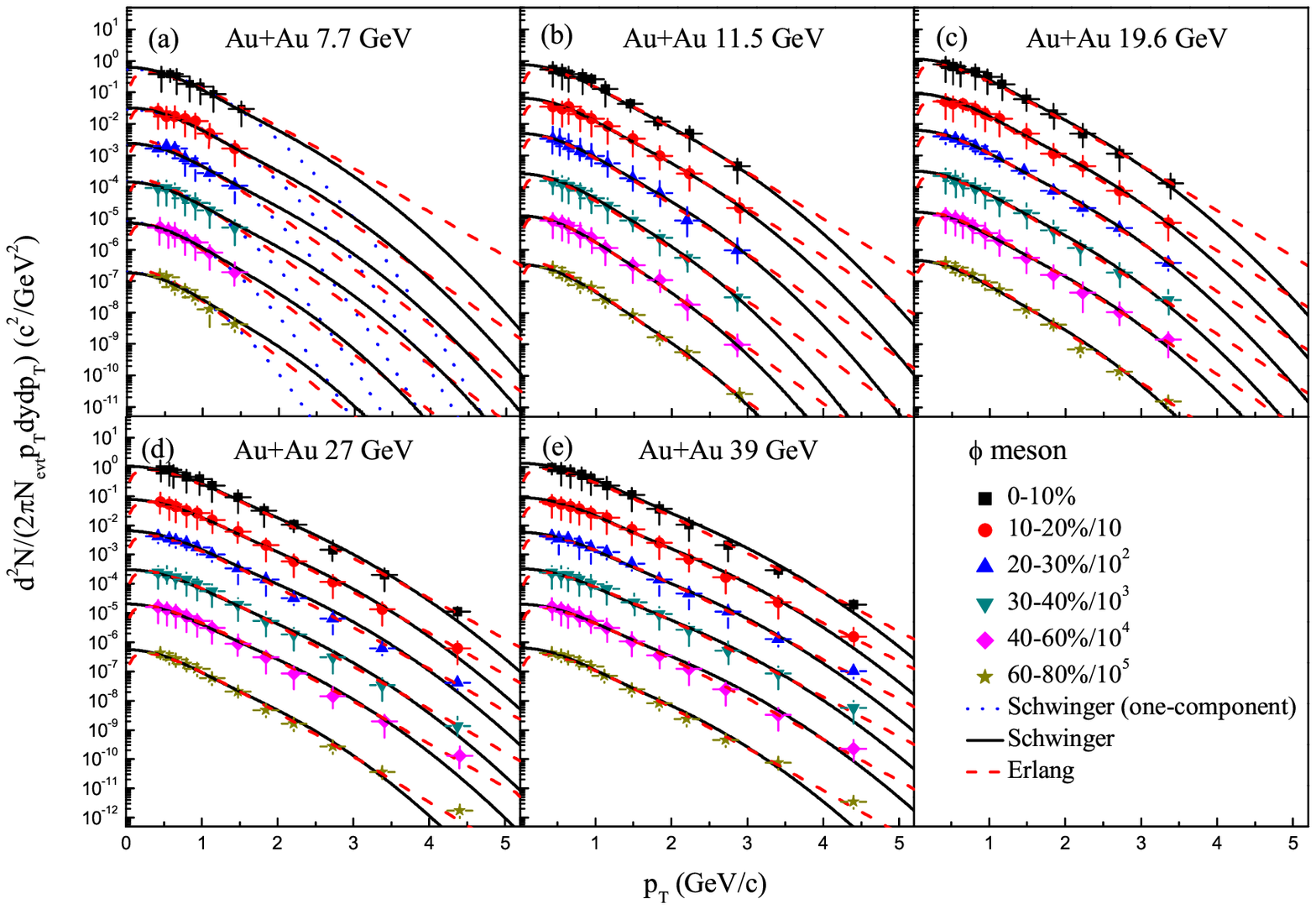}
\end{center}
\vskip.0cm Figure 1. Transverse momentum distributions of $\phi$
mesons produced in $|y|<0.5$ in Au-Au collisions at five RHIC BES
energies: (a) 7.7, (b) 11.5, (c) 19.6, (d) 27, and (e) 39 GeV. The
symbols represent the experimental data of the STAR Collaboration
[35] in different centrality intervals, which are scaled by
different amounts marked in the panel. The dashed and solid curves
are our results calculated by using the single Erlang distribution
and the two-component Schwinger mechanism respectively. The dotted
curves in Figure 1(a) are an example calculated by using the
single (one-component) Schwinger mechanism.
\end{figure}

\renewcommand{\baselinestretch}{.4}
\newpage
{\small {Table 1. Values of parameters $k_{S1}$, $\kappa_1$,
$\kappa_2$, and $\langle p_t \rangle$, as well as $\chi^2$/dof
corresponding to the solid and dashed curves in Fig. 1. In all the
cases, for the single Erlang distribution, we have $n=3$ which are
not listed in the column.
{%
\begin{center}
\begin{tabular}{cccccccc}
\hline\hline & & \multicolumn{4}{c}{Two-component Schwinger mechanism} & \multicolumn{2}{c}{Single Erlang distribution} \\
\cline{3-6} \cline{7-8}
Figure & Type & $k_{S1}$ & $\kappa_1$ (GeV/fm) & $\kappa_2$ (GeV/fm) & $\chi^2$/dof & $\langle p_t \rangle$ (GeV/$c$) & $\chi^2$/dof \\
\hline
Figure 1(a) & 0--10\%  & $0.628\pm0.135$ & $0.967\pm0.132$ & $2.635\pm0.245$ & 0.050 & $0.287\pm0.057$ & 0.076 \\
            & 10--20\% & $0.620\pm0.133$ & $0.952\pm0.127$ & $2.598\pm0.228$ & 0.120 & $0.247\pm0.046$ & 0.160 \\
            & 20--30\% & $0.615\pm0.137$ & $0.936\pm0.133$ & $2.564\pm0.237$ & 0.225 & $0.231\pm0.043$ & 0.288 \\
            & 30--40\% & $0.612\pm0.132$ & $0.895\pm0.135$ & $2.473\pm0.238$ & 0.066 & $0.230\pm0.038$ & 0.185 \\
            & 40--60\% & $0.605\pm0.128$ & $0.862\pm0.132$ & $2.126\pm0.219$ & 0.066 & $0.228\pm0.040$ & 0.092 \\
            & 60--80\% & $0.596\pm0.133$ & $0.787\pm0.137$ & $1.985\pm0.225$ & 0.364 & $0.208\pm0.038$ & 0.116 \\
Figure 1(b) & 0--10\%  & $0.632\pm0.132$ & $0.988\pm0.306$ & $2.682\pm0.382$ & 0.269 & $0.273\pm0.037$ & 0.516 \\
            & 10--20\% & $0.624\pm0.124$ & $0.953\pm0.313$ & $2.500\pm0.375$ & 0.096 & $0.267\pm0.033$ & 0.183 \\
            & 20--30\% & $0.618\pm0.108$ & $0.922\pm0.298$ & $2.266\pm0.323$ & 0.113 & $0.247\pm0.033$ & 0.135 \\
            & 30--40\% & $0.613\pm0.115$ & $0.892\pm0.303$ & $2.000\pm0.337$ & 0.020 & $0.238\pm0.035$ & 0.152 \\
            & 40--60\% & $0.600\pm0.113$ & $0.870\pm0.312$ & $1.892\pm0.357$ & 0.063 & $0.228\pm0.032$ & 0.200 \\
            & 60--80\% & $0.588\pm0.122$ & $0.823\pm0.297$ & $1.870\pm0.345$ & 0.214 & $0.218\pm0.035$ & 0.375 \\
Figure 1(c) & 0--10\%  & $0.635\pm0.157$ & $1.092\pm0.272$ & $2.735\pm0.336$ & 0.112 & $0.278\pm0.042$ & 0.186 \\
            & 10--20\% & $0.626\pm0.148$ & $0.976\pm0.304$ & $2.642\pm0.358$ & 0.115 & $0.273\pm0.036$ & 0.525 \\
            & 20--30\% & $0.610\pm0.153$ & $0.952\pm0.285$ & $2.553\pm0.326$ & 0.129 & $0.262\pm0.032$ & 0.093 \\
            & 30--40\% & $0.602\pm0.155$ & $0.938\pm0.287$ & $2.493\pm0.347$ & 0.132 & $0.260\pm0.035$ & 0.083 \\
            & 40--60\% & $0.593\pm0.157$ & $0.923\pm0.257$ & $2.456\pm0.334$ & 0.967 & $0.258\pm0.033$ & 0.121 \\
            & 60--80\% & $0.590\pm0.152$ & $0.832\pm0.232$ & $2.180\pm0.305$ & 0.364 & $0.242\pm0.038$ & 0.558 \\
Figure 1(d) & 0--10\%  & $0.642\pm0.162$ & $1.238\pm0.233$ & $3.266\pm0.256$ & 0.703 & $0.290\pm0.046$ & 0.428 \\
            & 10--20\% & $0.639\pm0.177$ & $1.182\pm0.202$ & $3.188\pm0.233$ & 0.422 & $0.286\pm0.040$ & 0.130 \\
            & 20--30\% & $0.626\pm0.172$ & $1.128\pm0.183$ & $3.105\pm0.225$ & 1.278 & $0.282\pm0.036$ & 0.398 \\
            & 30--40\% & $0.617\pm0.168$ & $1.080\pm0.184$ & $2.892\pm0.217$ & 0.475 & $0.275\pm0.040$ & 0.111 \\
            & 40--60\% & $0.592\pm0.176$ & $0.993\pm0.193$ & $2.856\pm0.206$ & 1.493 & $0.275\pm0.035$ & 0.390 \\
            & 60--80\% & $0.588\pm0.188$ & $0.870\pm0.170$ & $2.477\pm0.200$ & 9.734 & $0.258\pm0.038$ & 1.112 \\
Figure 1(e) & 0--10\%  & $0.640\pm0.186$ & $1.278\pm0.233$ & $3.463\pm0.313$ & 0.685 & $0.300\pm0.030$ & 0.223 \\
            & 10--20\% & $0.635\pm0.182$ & $1.182\pm0.205$ & $3.448\pm0.322$ & 0.427 & $0.308\pm0.028$ & 0.322 \\
            & 20--30\% & $0.628\pm0.173$ & $1.160\pm0.212$ & $3.396\pm0.302$ & 1.151 & $0.300\pm0.032$ & 0.385 \\
            & 30--40\% & $0.613\pm0.178$ & $1.129\pm0.189$ & $3.365\pm0.295$ & 0.756 & $0.298\pm0.028$ & 0.213 \\
            & 40--60\% & $0.608\pm0.180$ & $1.098\pm0.185$ & $3.242\pm0.290$ & 0.398 & $0.285\pm0.030$ & 0.242 \\
            & 60--80\% & $0.596\pm0.177$ & $0.905\pm0.190$ & $2.750\pm0.293$ & 4.921 & $0.277\pm0.027$ & 0.842 \\
\hline
\hline
\end{tabular}%
\end{center}
\renewcommand{\baselinestretch}{1.}

\begin{figure}
\hskip-1.0cm \begin{center}
\includegraphics[width=11.0cm]{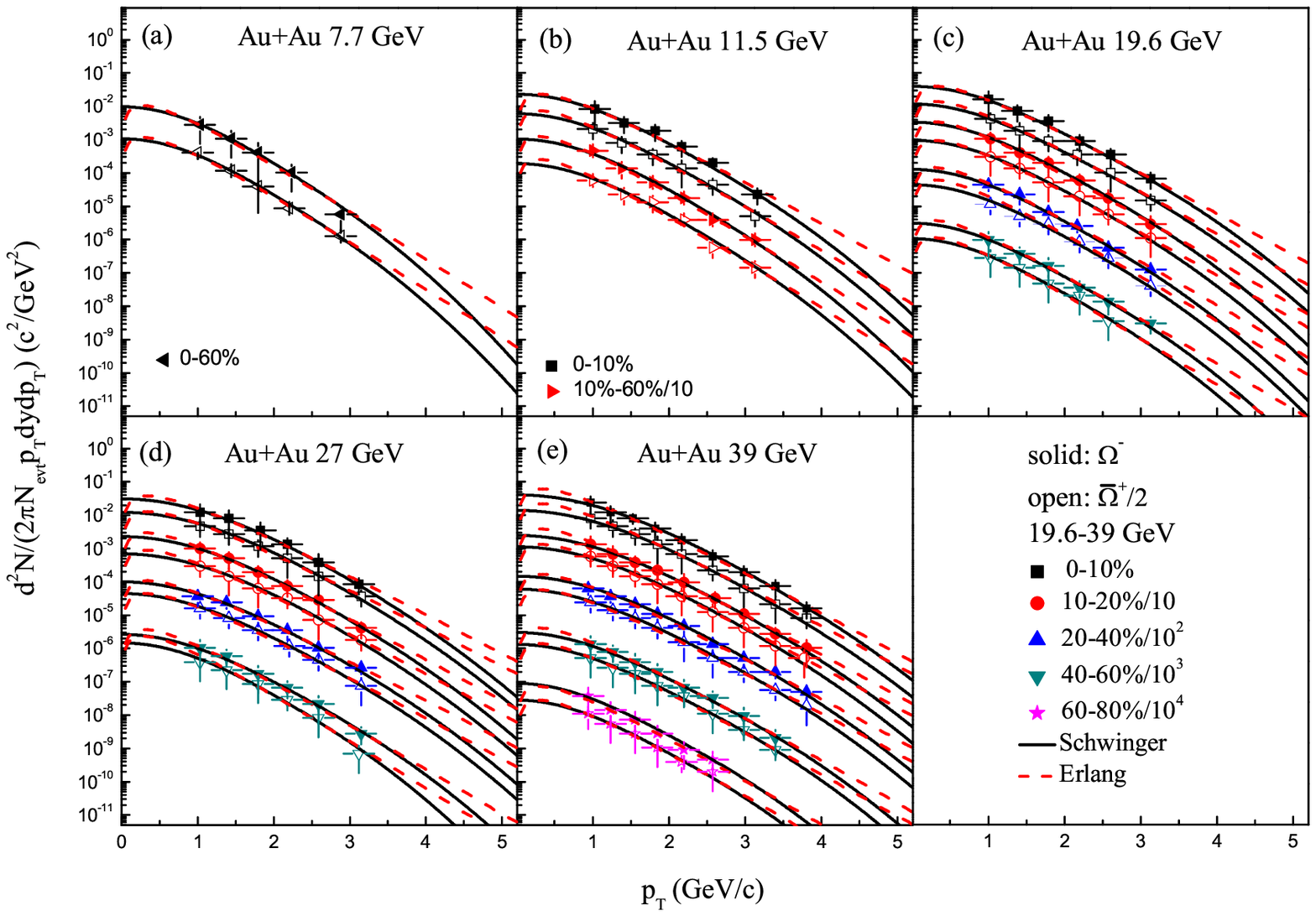}
\end{center}
\vskip.0cm Figure 2. The same as for Fig. 1, but showing the
results of $\Omega^-$ (solid symbols) and $\bar \Omega^+$ (open
symbols divided by 2) in different centrality intervals, where the
symbols represent the data of the STAR Collaboration [35]. At the
same time, the results of the one-component Schwinger mechanism
are not available.
\end{figure}

\renewcommand{\baselinestretch}{0.4}
\newpage
{\small {Table 2. Values of parameters $k_{S1}$, $\kappa_1$,
$\kappa_2$, and $\langle p_t \rangle$, as well as $\chi^2$/dof
corresponding to the solid and dashed curves in Fig. 2. In all the
cases, for the single Erlang distribution, we have $n=3$ which are
not listed in the column.
{%
\begin{center}
\begin{tabular}{cccccccc}
\hline\hline
& & \multicolumn{4}{c}{Two-component Schwinger mechanism} & \multicolumn{2}{c}{Single Erlang distribution} \\
\cline{3-6} \cline{7-8}
Figure & Type & $k_{S1}$ & $\kappa_1$ (GeV/fm) & $\kappa_2$ (GeV/fm) & $\chi^2$/dof & $\langle p_t \rangle$ (GeV/$c$) & $\chi^2$/dof \\
\hline
Figure 2(a) & 0--60\%  & $0.652\pm0.118$ & $1.564\pm0.388$ & $2.870\pm0.298$ & 0.145 & $0.280\pm0.026$ & 0.712 \\
            & 0--60\%  & $0.652\pm0.103$ & $1.580\pm0.350$ & $2.912\pm0.303$ & 0.270 & $0.282\pm0.022$ & 0.828 \\
Figure 2(b) & 0--10\%  & $0.635\pm0.115$ & $1.788\pm0.373$ & $3.432\pm0.305$ & 0.304 & $0.321\pm0.031$ & 0.446 \\
            & 0--10\%  & $0.635\pm0.103$ & $1.782\pm0.315$ & $3.420\pm0.300$ & 0.104 & $0.302\pm0.028$ & 0.288 \\
            & 10--60\% & $0.632\pm0.126$ & $1.753\pm0.310$ & $3.213\pm0.288$ & 0.209 & $0.293\pm0.032$ & 0.291 \\
            & 10--60\% & $0.630\pm0.118$ & $1.732\pm0.272$ & $3.202\pm0.282$ & 0.329 & $0.288\pm0.030$ & 0.659 \\
Figure 2(c) & 0--10\%  & $0.642\pm0.120$ & $1.852\pm0.332$ & $3.725\pm0.332$ & 0.124 & $0.325\pm0.035$ & 0.264 \\
            & 0--10\%  & $0.635\pm0.117$ & $1.786\pm0.327$ & $3.693\pm0.318$ & 0.134 & $0.315\pm0.027$ & 0.286 \\
            & 10--20\% & $0.637\pm0.122$ & $1.715\pm0.303$ & $3.387\pm0.302$ & 0.042 & $0.303\pm0.030$ & 0.191 \\
            & 10--20\% & $0.630\pm0.118$ & $1.706\pm0.308$ & $3.372\pm0.307$ & 0.037 & $0.310\pm0.032$ & 0.068 \\
            & 20--40\% & $0.627\pm0.124$ & $1.658\pm0.312$ & $3.243\pm0.297$ & 0.109 & $0.306\pm0.034$ & 0.100 \\
            & 20--40\% & $0.613\pm0.113$ & $1.643\pm0.298$ & $3.224\pm0.290$ & 0.186 & $0.300\pm0.032$ & 0.557 \\
            & 40--60\% & $0.610\pm0.124$ & $1.543\pm0.292$ & $3.178\pm0.288$ & 0.091 & $0.296\pm0.028$ & 0.128 \\
            & 40--60\% & $0.602\pm0.118$ & $1.482\pm0.295$ & $3.076\pm0.284$ & 0.121 & $0.290\pm0.030$ & 0.295 \\
Figure 2(d) & 0--10\%  & $0.657\pm0.108$ & $2.420\pm0.372$ & $3.878\pm0.382$ & 0.049 & $0.343\pm0.028$ & 0.215 \\
            & 0--10\%  & $0.635\pm0.102$ & $2.395\pm0.398$ & $3.876\pm0.400$ & 0.035 & $0.349\pm0.035$ & 0.117 \\
            & 10--20\% & $0.644\pm0.111$ & $2.246\pm0.364$ & $3.772\pm0.362$ & 0.041 & $0.326\pm0.030$ & 0.218 \\
            & 10--20\% & $0.630\pm0.114$ & $2.356\pm0.352$ & $3.770\pm0.380$ & 0.054 & $0.332\pm0.032$ & 0.236 \\
            & 20--40\% & $0.612\pm0.103$ & $2.176\pm0.357$ & $3.757\pm0.337$ & 0.097 & $0.340\pm0.038$ & 0.210 \\
            & 20--40\% & $0.607\pm0.100$ & $2.030\pm0.360$ & $3.684\pm0.364$ & 0.026 & $0.330\pm0.030$ & 0.171 \\
            & 40--60\% & $0.598\pm0.123$ & $1.843\pm0.352$ & $3.472\pm0.353$ & 0.133 & $0.302\pm0.036$ & 0.317 \\
            & 40--60\% & $0.590\pm0.102$ & $1.662\pm0.346$ & $3.010\pm0.230$ & 0.179 & $0.282\pm0.025$ & 0.610 \\
Figure 2(e) & 0--10\%  & $0.678\pm0.120$ & $2.564\pm0.324$ & $4.235\pm0.325$ & 0.066 & $0.331\pm0.026$ & 0.152 \\
            & 0--10\%  & $0.668\pm0.124$ & $2.478\pm0.328$ & $4.192\pm0.302$ & 0.128 & $0.335\pm0.028$ & 0.087 \\
            & 10--20\% & $0.653\pm0.115$ & $2.346\pm0.336$ & $4.056\pm0.316$ & 0.061 & $0.328\pm0.032$ & 0.180 \\
            & 10--20\% & $0.647\pm0.117$ & $2.260\pm0.320$ & $3.928\pm0.318$ & 0.199 & $0.333\pm0.027$ & 0.080 \\
            & 20--40\% & $0.640\pm0.108$ & $2.232\pm0.318$ & $3.937\pm0.303$ & 0.220 & $0.330\pm0.025$ & 0.108 \\
            & 20--40\% & $0.626\pm0.117$ & $2.030\pm0.322$ & $3.883\pm0.322$ & 0.077 & $0.332\pm0.027$ & 0.039 \\
            & 40--60\% & $0.615\pm0.121$ & $1.856\pm0.326$ & $3.764\pm0.315$ & 0.074 & $0.315\pm0.035$ & 0.209 \\
            & 40--60\% & $0.602\pm0.126$ & $1.802\pm0.298$ & $3.650\pm0.308$ & 0.047 & $0.318\pm0.032$ & 0.159 \\
            & 60--80\% & $0.587\pm0.127$ & $1.588\pm0.292$ & $3.202\pm0.313$ & 0.266 & $0.305\pm0.025$ & 0.271 \\
            & 60--80\% & $0.580\pm0.118$ & $1.526\pm0.302$ & $3.110\pm0.310$ & 0.118 & $0.302\pm0.028$ & 0.224 \\
\hline \hline
\end{tabular}%
\end{center}
}} }
\renewcommand{\baselinestretch}{1.}

\begin{figure}
\hskip-1.0cm \begin{center}
\includegraphics[width=11.0cm]{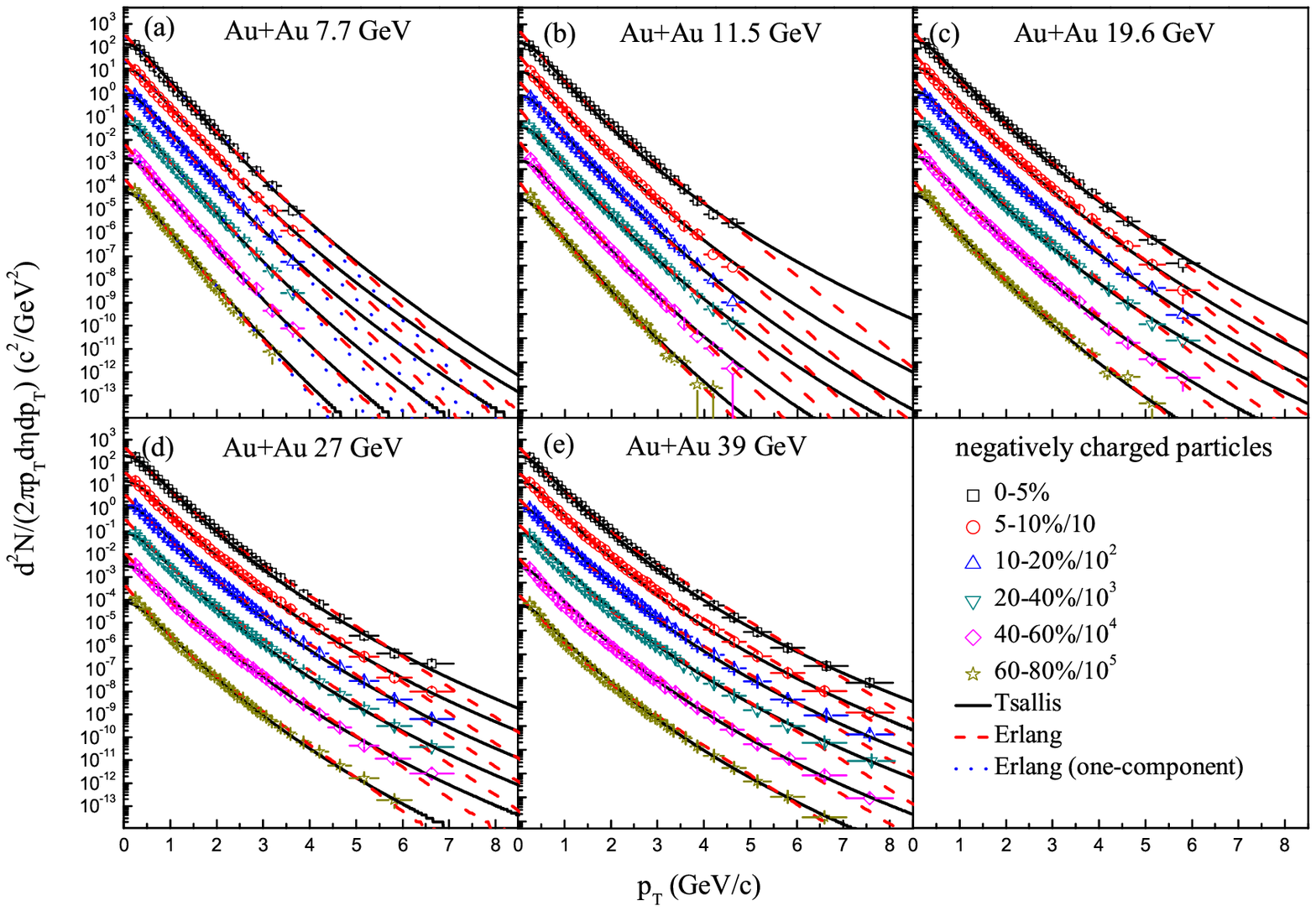}
\end{center}
\vskip.0cm Figure 3. The same as for Fig. 1, but showing the
results of negatively charged particles in $|\eta|<0.5$ in
different centrality intervals, where the symbols represent the
data of the STAR Collaboration [36], and the dashed and solid
curves are our results calculated by using the two-component
Erlang distribution and the single Tsallis statistics
respectively. The dotted curves in Figure 3(a) are an example
calculated by using the one-component Erlang distribution.
\end{figure}

\renewcommand{\baselinestretch}{.4}
\newpage
{\small {Table 3. Values of parameters $T$, $q$, $k_{E1}$,
$\langle p_t \rangle_1$, and $\langle p_t \rangle_2$, as well as
$\chi^2$/dof corresponding to the solid and dashed curves in Fig.
3. In all the cases, for the two-component Erlang distribution, we
have $n_1=3$ and $n_2=2$ which are not listed in the column.
{%
\begin{center}
\begin{tabular}{ccccccccc}
\hline\hline
& & \multicolumn{3}{c}{Single Tsallis statistics} & \multicolumn{4}{c}{Two-component Erlang distribution} \\
\cline{3-5} \cline{6-9}
Figure & Type & $T$ (GeV) & $q$ & $\chi^2$/dof & $k_{E1}$ & $\langle p_t \rangle_1$ (GeV/$c$) & $\langle p_t \rangle_2$ (GeV/$c$) & $\chi^2$/dof \\
\hline
Figure 3(a) & 0--5\%   & $0.152\pm0.014$ & $1.024\pm0.008$ & 1.442 & $0.821\pm0.019$ & $0.189\pm0.013$ & $0.240\pm0.024$ & 0.886 \\
            & 5--10\%  & $0.152\pm0.012$ & $1.026\pm0.007$ & 1.059 & $0.816\pm0.021$ & $0.187\pm0.017$ & $0.238\pm0.026$ & 2.648 \\
            & 10--20\% & $0.151\pm0.013$ & $1.025\pm0.008$ & 1.924 & $0.812\pm0.017$ & $0.189\pm0.014$ & $0.232\pm0.024$ & 2.156 \\
            & 20--40\% & $0.151\pm0.015$ & $1.022\pm0.006$ & 1.825 & $0.810\pm0.017$ & $0.183\pm0.013$ & $0.224\pm0.020$ & 0.860 \\
            & 40--60\% & $0.150\pm0.013$ & $1.020\pm0.008$ & 1.487 & $0.807\pm0.023$ & $0.180\pm0.010$ & $0.220\pm0.018$ & 1.585 \\
            & 60--80\% & $0.148\pm0.010$ & $1.018\pm0.008$ & 2.842 & $0.803\pm0.020$ & $0.176\pm0.012$ & $0.205\pm0.020$ & 1.634 \\
Figure 3(b) & 0--5\%   & $0.150\pm0.012$ & $1.043\pm0.006$ & 5.003 & $0.826\pm0.024$ & $0.188\pm0.016$ & $0.285\pm0.019$ & 2.199 \\
            & 5--10\%  & $0.152\pm0.014$ & $1.034\pm0.008$ & 3.678 & $0.824\pm0.027$ & $0.178\pm0.013$ & $0.258\pm0.022$ & 1.657 \\
            & 10--20\% & $0.150\pm0.011$ & $1.031\pm0.008$ & 4.948 & $0.817\pm0.026$ & $0.178\pm0.015$ & $0.246\pm0.018$ & 4.594 \\
            & 20--40\% & $0.150\pm0.012$ & $1.033\pm0.005$ & 2.410 & $0.795\pm0.022$ & $0.173\pm0.011$ & $0.245\pm0.013$ & 2.956 \\
            & 40--60\% & $0.150\pm0.014$ & $1.030\pm0.007$ & 2.983 & $0.786\pm0.024$ & $0.168\pm0.015$ & $0.234\pm0.017$ & 2.460 \\
            & 60--80\% & $0.148\pm0.008$ & $1.027\pm0.008$ & 4.885 & $0.773\pm0.023$ & $0.163\pm0.013$ & $0.225\pm0.016$ & 4.242 \\
Figure 3(c) & 0--5\%   & $0.152\pm0.010$ & $1.038\pm0.010$ & 4.495 & $0.835\pm0.027$ & $0.196\pm0.018$ & $0.297\pm0.023$ & 1.840 \\
            & 5--10\%  & $0.156\pm0.012$ & $1.035\pm0.006$ & 5.609 & $0.826\pm0.023$ & $0.187\pm0.013$ & $0.288\pm0.017$ & 1.180 \\
            & 10--20\% & $0.153\pm0.010$ & $1.036\pm0.007$ & 8.570 & $0.819\pm0.025$ & $0.183\pm0.016$ & $0.285\pm0.019$ & 2.184 \\
            & 20--40\% & $0.153\pm0.008$ & $1.039\pm0.006$ & 3.641 & $0.808\pm0.018$ & $0.178\pm0.020$ & $0.285\pm0.023$ & 3.866 \\
            & 40--60\% & $0.151\pm0.013$ & $1.037\pm0.008$ & 3.354 & $0.795\pm0.022$ & $0.173\pm0.014$ & $0.280\pm0.018$ & 3.457 \\
            & 60--80\% & $0.150\pm0.011$ & $1.034\pm0.006$ & 5.860 & $0.793\pm0.024$ & $0.168\pm0.017$ & $0.272\pm0.020$ & 2.821 \\
Figure 3(d) & 0--5\%   & $0.152\pm0.012$ & $1.043\pm0.006$ & 7.770 & $0.875\pm0.032$ & $0.213\pm0.018$ & $0.328\pm0.028$ & 3.921 \\
            & 5--10\%  & $0.154\pm0.014$ & $1.043\pm0.008$ & 6.243 & $0.870\pm0.028$ & $0.206\pm0.015$ & $0.326\pm0.024$ & 3.135 \\
            & 10--20\% & $0.154\pm0.011$ & $1.042\pm0.007$ & 5.889 & $0.866\pm0.026$ & $0.200\pm0.022$ & $0.325\pm0.025$ & 2.309 \\
            & 20--40\% & $0.155\pm0.012$ & $1.042\pm0.006$ & 3.860 & $0.862\pm0.030$ & $0.192\pm0.017$ & $0.320\pm0.027$ & 8.382 \\
            & 40--60\% & $0.155\pm0.013$ & $1.043\pm0.010$ & 3.818 & $0.858\pm0.030$ & $0.185\pm0.017$ & $0.320\pm0.024$ & 4.309 \\
            & 60--80\% & $0.148\pm0.012$ & $1.040\pm0.007$ & 5.483 & $0.853\pm0.027$ & $0.178\pm0.013$ & $0.300\pm0.022$ & 5.546 \\
Figure 3(e) & 0--5\%   & $0.150\pm0.013$ & $1.050\pm0.007$ & 3.193 & $0.920\pm0.035$ & $0.226\pm0.020$ & $0.370\pm0.024$ & 6.894 \\
            & 5--10\%  & $0.152\pm0.014$ & $1.048\pm0.006$ & 4.982 & $0.913\pm0.027$ & $0.223\pm0.017$ & $0.368\pm0.027$ & 3.643 \\
            & 10--20\% & $0.153\pm0.012$ & $1.048\pm0.007$ & 4.178 & $0.910\pm0.032$ & $0.218\pm0.015$ & $0.365\pm0.022$ & 5.659 \\
            & 20--40\% & $0.153\pm0.014$ & $1.049\pm0.009$ & 9.191 & $0.908\pm0.026$ & $0.216\pm0.017$ & $0.358\pm0.025$ & 9.063 \\
            & 40--60\% & $0.151\pm0.012$ & $1.049\pm0.006$ & 6.714 & $0.902\pm0.028$ & $0.210\pm0.014$ & $0.356\pm0.023$ & 6.465 \\
            & 60--80\% & $0.150\pm0.014$ & $1.049\pm0.008$ & 4.322 & $0.895\pm0.025$ & $0.200\pm0.015$ & $0.352\pm0.024$ & 6.094 \\
\hline \hline
\end{tabular}%
\end{center}
}} }
\renewcommand{\baselinestretch}{1.}

To give a comparison, Figure 3 gives the results, $d^2N/(2\pi p_T
d\eta dp_T)$, of negatively charged particles (hadrons) ($h^-$)
produced in mid-pseudorapidity ($|\eta|<0.5$) in the same
collisions with a bit difference in centrality intervals, where
$h^-$ are assumed to produce mostly in soft process and the
contribution of hard process to $h^-$ with high $p_T$ is neglected
due to small amount. The symbols represent the data of the STAR
Collaboration [36], and the dashed and solid curves are our
results calculated by using the two-component Erlang distribution
and the single Tsallis statistics respectively. The dotted curves
in Figure 3(a) are the results of single Erlang distribution,
which are given as an example and are not in agreement with the
spectra in wide transverse momentum range. The values of $n_1$ and
$n_2$ for all the cases are taken to be 3 and 2 respectively.
Other related parameter values related to the two-component Erlang
distribution and the single Tsallis statistics are listed in Table
3 with values of $\chi^2$/dof. The chemical potentials related to
collision energy [37] in Eq. (5) are taken to be 0.407, 0.304,
0.197, 0.149, and 0.107 GeV which correspond to 7.7, 11.5, 19.6,
27, and 39 GeV, respectively. For the dotted curves in Figure 3(a)
and the centralities from 0--5\% to 60--80\%, the values of
$\langle p_t \rangle$ in the single Erlang distribution are taken
to be $0.212\pm0.025$, $0.210\pm0.023$, $0.209\pm0.022$,
$0.203\pm0.018$, $0.194\pm0.021$, and $0.186\pm0.022$ GeV/$c$, and
the values of $n$ for all the six cases are taken to be 3, which
result in the values of $\chi^2$/dof to be 0.922, 2.430, 1.980,
0.930, 1.374, and 1.965, respectively. One can see that the
two-component Erlang distribution and the single Tsallis
statistics describe approximately the experimental data of $h^-$
produced in mid-pseudorapidity in Au-Au collisions at the RHIC BES
in most cases.

To study the change trends of parameters with centrality interval
($C$) and collision energy ($\sqrt{s_{NN}}$), Figure 4 shows the
dependences of parameters $k_{S1}$, $\kappa_1$, $\kappa_2$, and
$\langle p_t \rangle$ on centrality at different energies (left
panel) and on energy in different centrality intervals (right
panel). The different symbols represent the parameter values
extracted from Figure 1 and listed in Table 1, where the
two-component Schwinger mechanism and single Erlang distribution
are used for $\phi$ spectra. The lines are our fitted results, and
the values of intercepts, slopes, and $\chi^2$/dof are listed in
Table 4. Figure 5 is the same as that for Figure 4, but it shows
the parameter values extracted from Figure 2 and listed in Table
2, where the two-component Schwinger mechanism and single Erlang
distribution are used for $\Omega^-$ (solid symbols) and
$\bar\Omega^+$ (open symbols) spectra respectively. In particular,
the solid and open symbols in Figure 5 are shifted respectively to
the left and right sides by a small amount for clearness. The
values of intercepts, slopes, and $\chi^2$/dof corresponding to
the lines in Figure 5 are listed in Table 5. Figure 6 displays the
dependences of parameters $T$, $q$, $k_{E1}$, $\langle p_t
\rangle_1$, and $\langle p_t \rangle_2$ on centrality at different
energies (left panel) and on energy in different centrality
intervals (right panel). The different symbols represent the
parameter values extracted from Figure 3 and listed in Table 3,
where the single Tsallis statistics and two-component Erlang
distribution are used for negatively charged particle spectra. The
values of intercepts, slopes, and $\chi^2$/dof corresponding to
the lines in Figure 6 are listed in Table 6.

One can see from Figures 4--6 that the relative contribution of
the first component in the two-component Schwinger mechanism
decreases slightly with decrease of centrality, and does not show
an obvious change with change of collision energy [Figures 4(a),
4(b), 5(a), and 5(b)]. The relative contribution of the first
component in the two-component Erlang distribution decreases
slightly with decrease of centrality, and increases with increase
of collision energy [Figures 6(e) and 6(f)]. The mean transverse
momentum contributed by each parton in the Erlang distribution
decrease with decrease of centrality [Figures 4(g), 5(g), 6(g),
and 6(i)], and increase with increase of collision energy [Figures
4(h), 5(h), 6(h), and 6(j)]. The string tension between partons in
the Schwinger mechanism decrease also with decrease of centrality
[Figures 4(c), 4(e), 5(c), and 5(e)], and increase also with
increase of collision energy [Figures 4(d), 4(f), 5(d), and 5(f)].
The effective temperature of emission source in the Tsallis
statistics does not show an obvious dependence on centrality and
collision energy [Figure 6(a) and 6(b)]. The entropy index does
not show an obvious change with change of centrality, and increase
with increase of collision energy [Figures 6(c) and 6(d)]. A weak
softest point in the string tension in $\Omega$ hyperon spectra is
observed at 7.7 GeV [Figures 5(e) and 5(g)].

To study the relations between different distributions, as well as
their parameters, we could give some further discussions. It
should be noted that Figures 1 and 2 devoted to multi-strange
hadrons $\phi(s\bar s)$ mesons and $\Omega(sss)$ hyperons
respectively. Both $\phi(s\bar s)$ mesons and $\Omega(sss)$
hyperons are expected to have relatively small hadronic
interaction cross-sections. They are also important probes for the
search of the QGP phase transition [35]. We have used the
two-component Schwinger mechanism and the single Erlang
distribution to describe the spectra of multi-strange hadrons
which are expected to be produced in hard scattering process.
Although the Schwinger mechanism can give the string tension
$\kappa$ between partons, the flexible Erlang distribution can
describe the violent degree of parton-parton interactions or
excitation degree of interacting system (emission source) by the
mean transverse momentum $\langle p_t \rangle$ contributed by each
parton and the mean transverse momentum $\langle p_T \rangle$
contributed by $n$ partons. From the comparisons between the
Schwinger mechanism and Erlang distribution in Figure 1 and Table
1, we can obtain the linear relations between $\kappa$ and
$\langle p_T \rangle$ for $\phi$ mesons in Figures 7(a) and 7(b)
which correspond respectively to different energies (in various
centrality intervals) and to different centrality intervals (at
various energies). Similarly, the results for $\Omega$ ($\Omega^-$
and $\bar\Omega^+$) hyperons are presented in Figures 7(c) and
7(d) based on Figure 2 and Table 2. Different symbols represent
different energies and centrality intervals, where $C_1$, $C_2$,
$\cdots$, $C_6$ for Figures 7(b) and 7(d) represent respectively
and successively the centrality intervals mentioned in the right
panels of Figures 4 and 5, and the lines are our fitted results.
The values of intercepts, slopes, and $\chi^2$/dof are listed in
Table 7. One can see that $\kappa$ increases with increase of
$\langle p_T \rangle$. There is a one-to-one correspondence
between the two-component Schwinger mechanism and the single
Erlang distribution.

For the spectra of negatively charged hadrons presented in Figure
3, we have used the single Tsallis statistics and the
two-component Erlang distribution. Because of the Tsallis
statistics resulting in the sum of two or three standard
distribution [38], it has the temperature parameter $T$ which
describes the excitation degree of interacting system. The mean
transverse momentum $\langle p_T \rangle$ obtained from the
two-component Erlang distribution also describes the excitation
degree of interacting system. From the comparisons between the
single Tsallis statistics and the two-component Erlang
distribution in Figure 3 and Table 3, we can obtain the linear
relations between $T$ and $\langle p_T \rangle$ in Figures 7(e)
and 7(f) which correspond respectively to different collision
energies (in various centrality intervals) and to different
centrality intervals (at various energies), where $C_1$, $C_2$,
$\cdots$, $C_6$ for Figure 7(f) represent successively the
centrality intervals mentioned in the right panel of Figure 6. The
values of intercepts, slopes, and $\chi^2$/dof are listed in Table
7. One can see that $T$ does not show an obvious change with
increase of $\langle p_T \rangle$. Instead, Figure 6 shows that
$q$ and $\langle p_T \rangle$ increase with increase of
$\sqrt{s_{NN}}$, which renders that $q$ increases with increase of
$\langle p_T \rangle$. There is a one-to-one correspondence
between the single Tsallis statistics and the two-component Erlang
distribution. Because the Schwinger mechanism and Tsallis
statistics do not describe the same set of experimental data. We
do not expect to study the relation between $\kappa$ and $T$. In
addition, in the case of considering the two-component functions,
we obtain the values of $\kappa$, $T$, and $\langle p_T \rangle$
discussed above to be respectively an average weighted by the two
components.

In our recent works [39, 40], we have studied the two-component
Schwinger mechanism and the two-component Erlang distribution for
$J/\psi$ and $\Upsilon$ productions [39], and L{\'e}vy
distribution and the multi-component Erlang distribution for
identified particles productions [40]. In most cases, different
functions (Schwinger and Eralng or Tsallis and Erlang) describe
the same set of experimental data (hard or soft particles), which
reflects the relations among them. The relations between $\kappa$
and $\langle p_T \rangle$, as well as $T$ and $\langle p_T
\rangle$, discussed above are an attempt to study these relations.
In particular, the single and two-component Erlang distributions
describe the particle spectra in hard and soft processes
respectively, which reflects the Erlang distribution being
flexible. We think that there are some universal laws existing in
the two processes. If the hard process corresponds to a violent
collision, the soft process corresponds to a non-violent or a very
non-violent collision. Only from the collision itself, both the
hard and soft processes obey the same Erlang distribution with
different $\langle p_t \rangle$ and parton numbers. We think that
the Erlang distribution is one of the universal laws existing in
the hard and soft processes.
\\

{\section{Conclusions}}

We summarize here our main observations and conclusions.

a) The transverse momentum distributions of $\phi$ mesons,
$\Omega$ hyperons, and negatively charged particles produced in
mid-(pseudo)rapidity in Au-Au collisions with different centrality
intervals at RHIC BES energies are analyzed by using the single or
two-component Erlang distribution, the two-component Schwinger
mechanism, and the single Tsallis statistics. The single or
two-component Erlang distribution is approximately in agreement
with the experimental data measured by the STAR Collaboration over
an energy range from 7.7 to 39 GeV. The two-component Schwinger
mechanism describes approximately the data of $\phi$ mesons and
$\Omega$ hyperons which are assumed to produce in hard process.
The single Tsallis statistics describes approximately the data of
negatively charged particles in which most are assumed to produce
in soft process.

b) The mean transverse momentum contributed by each parton in the
Erlang distribution which describes $\phi$ mesons, $\Omega$
hypersons, and negatively charged particles, and the string
tension between partons in the Schwinger mechanism which describes
$\phi$ mesons and $\Omega$ hypersons, decrease with decrease of
centrality, and increase with increase of collision energy. The
effective temperature of emission source in the Tsallis statistics
which describes the negatively charged particles does not show an
obvious dependence on centrality and collision energy. The entropy
index does not show an obvious change with change of centrality,
and increase with increase of collision energy. A weak softest
point in the string tension in $\Omega$ hyperon spectra is
observed at 7.7 GeV.

c) The relative contribution of the first component in the
two-component Erlang distribution which describes the negatively
charged particles decreases slightly with decrease of centrality,
and increases with increase of collision energy. The relative
contribution of the first component in the two-component Schwinger
mechanism which describes $\phi$ mesons and $\Omega$ hypersons
decreases slightly with decrease of centrality, and does not show
an obvious change with change of collision energy.

d) In the descriptions of hard process, the string tension in the
two-component Schwinger mechanism increases with increase of the
mean transverse momentum in the single Erlang distribution. There
is a one-to-one correspondence between the two descriptions for
hard process. In the descriptions of soft process, the temperature
and entropy index in the single Tsallis statistics unchanges
approximately or increases with increase of mean transverse
momentum in the two-component Erlang distribution. There is also a
one-to-one correspondence between the two descriptions for soft
process.
\\
\\
\\
\\
{\bf Conflict of interests}

The authors declare that there is no conflict of interests
regarding the publication of this paper.
\\
\\
\\
{\bf Acknowledgments}

This work was supported by the National Natural Science Foundation
of China under Grant No. 11575103 and the US DOE under contract
DE-FG02-87ER40331.A008.
\\

\begin{figure}
\hskip-1.0cm \begin{center}
\includegraphics[width=15.0cm]{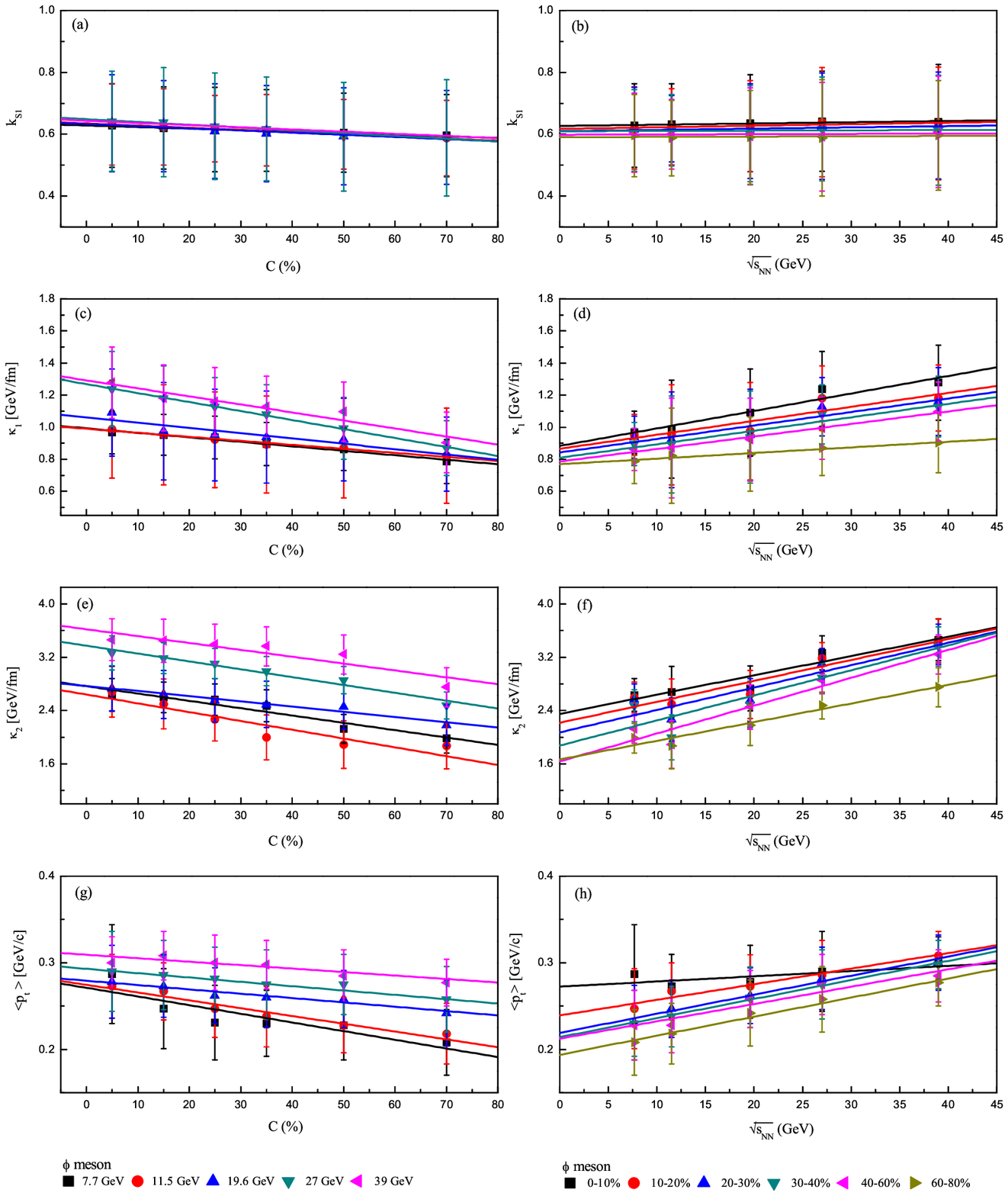}
\end{center}
\vskip.0cm Figure 4. Dependences of parameters (a,b) $k_{S1}$,
(c,d) $\kappa_1$, (e,f) $\kappa_2$, and (g,h) $\langle p_t
\rangle$ on centrality at different energies (left panel) and on
energy in different centrality intervals (right panel). The
different symbols represent the parameter values extracted from
Fig. 1 and listed in Table 1, where the two-component Schwinger
mechanism and single Erlang distribution are used for $\phi$
spectra. The lines are our fitted results.
\end{figure}

\renewcommand{\baselinestretch}{0.4}
\newpage
{\small {Table 4. Values of intercepts, slopes, and $\chi^2$/dof
corresponding to the lines in Fig. 4 which shows dependences of
parameters on centrality at different energies (a,c,e,g) and on
energy in different centrality intervals (b,d,f,h). The values of
parameters are extracted from Fig. 1 and listed in Table 1.
{%
\begin{center}
\begin{tabular}{cccccc}
\hline\hline  Figure & Relation & Type & Intercept & Slope & $\chi^2$/dof \\
\hline
& & & \multicolumn{3}{c}{Two-component Schwinger mechanism in Fig. 1} \\
\cline{4-6}
Figure 4(a) & $k_{S1}-C$ & 7.7 GeV  & $0.628\pm0.001$ & $-0.00047\pm0.00002$ & 0.001 \\
            && 11.5 GeV & $0.635\pm0.001$ & $-0.00067\pm0.00002$ & 0.001 \\
            && 19.6 GeV & $0.633\pm0.004$ & $-0.00072\pm0.00011$ & 0.002 \\
            && 27 GeV   & $0.649\pm0.004$ & $-0.00094\pm0.00010$ & 0.001 \\
            && 39 GeV   & $0.644\pm0.001$ & $-0.00070\pm0.00006$ & 0.001 \\
Figure 4(b) & $k_{S1}-\sqrt{s_{NN}}$ & 0--10\%  & $0.627\pm0.002$ & $0.00040\pm0.00011$  & 0.001 \\
            && 10--20\% & $0.618\pm0.004$ & $0.00054\pm0.00016$  & 0.001 \\
            && 20--30\% & $0.610\pm0.005$ & $0.00044\pm0.00020$  & 0.002 \\
            && 30--40\% & $0.610\pm0.005$ & $0.00008\pm0.00022$  & 0.002 \\
            && 40--60\% & $0.598\pm0.007$ & $0.00006\pm0.00028$  & 0.003 \\
            && 60--80\% & $0.590\pm0.004$ & $0.00006\pm0.00016$  & 0.001 \\
Figure 4(c) & $\kappa_1-C$ & 7.7 GeV  & $0.993\pm0.007$ & $-0.00280\pm0.00019$ & 0.007 \\
            && 11.5 GeV & $0.990\pm0.006$ & $-0.00246\pm0.00015$ & 0.001 \\
            && 19.6 GeV & $1.062\pm0.024$ & $-0.00329\pm0.00059$ & 0.020 \\
            && 27 GeV   & $1.269\pm0.004$ & $-0.00561\pm0.00010$ & 0.001 \\
            && 39 GeV   & $1.292\pm0.026$ & $-0.00500\pm0.00065$ & 0.042 \\
Figure 4(d) & $\kappa_1-\sqrt{s_{NN}}$ & 0--10\%  & $0.883\pm0.034$ & $0.01093\pm0.00141$  & 0.030 \\
            && 10--20\% & $0.866\pm0.050$ & $0.00874\pm0.00212$  & 0.079 \\
            && 20--30\% & $0.843\pm0.042$ & $0.00844\pm0.00175$  & 0.063 \\
            && 30--40\% & $0.810\pm0.030$ & $0.00844\pm0.00124$  & 0.032 \\
            && 40--60\% & $0.786\pm0.012$ & $0.00777\pm0.00049$  & 0.007 \\
            && 60--80\% & $0.769\pm0.009$ & $0.00354\pm0.00037$  & 0.003 \\
Figure 4(e) & $\kappa_2-C$ & 7.7 GeV  & $2.640\pm0.095$ & $-0.01105\pm0.00140$ & 0.127 \\
            && 11.5 GeV & $2.771\pm0.030$ & $-0.01316\pm0.00238$ & 0.165 \\
            && 19.6 GeV & $3.374\pm0.040$ & $-0.00783\pm0.00077$ & 0.020 \\
            && 27 GeV   & $3.620\pm0.085$ & $-0.01180\pm0.00101$ & 0.080 \\
            && 39 GeV   & $2.765\pm0.056$ & $-0.01029\pm0.00214$ & 0.185 \\
Figure 4(f) & $\kappa_2-\sqrt{s_{NN}}$ & 0--10\%  & $2.351\pm0.112$ & $0.02887\pm0.00469$  & 0.213 \\
            && 10--20\% & $2.217\pm0.132$ & $0.03142\pm0.00553$  & 0.326 \\
            && 20--30\% & $2.073\pm0.177$ & $0.03357\pm0.00743$  & 0.643 \\
            && 30--40\% & $1.876\pm0.220$ & $0.03763\pm0.00924$  & 0.948 \\
            && 40--60\% & $1.638\pm0.140$ & $0.04182\pm0.00590$  & 0.391 \\
            && 60--80\% & $1.666\pm0.081$ & $0.02797\pm0.00339$  & 0.140 \\
\hline
& & & \multicolumn{3}{c}{Single Erlang distribution in Fig. 1} \\
\cline{4-6}
Figure 4(g) & $\langle p_t \rangle-C$ & 7.7 GeV  & $0.271\pm0.010$ & $-0.00098\pm0.00025$ & 0.095 \\
            && 11.5 GeV & $0.275\pm0.004$ & $-0.00088\pm0.00010$ & 0.029 \\
            && 19.6 GeV & $0.279\pm0.002$ & $-0.00051\pm0.00006$ & 0.011 \\
            && 27 GeV   & $0.293\pm0.002$ & $-0.00046\pm0.00005$ & 0.007 \\
            && 39 GeV   & $0.309\pm0.003$ & $-0.00044\pm0.00009$ & 0.031 \\
Figure 4(h) & $\langle p_t \rangle-\sqrt{s_{NN}}$ & 0--10\%  & $0.272\pm0.007$ & $0.00063\pm0.00028$  & 0.032 \\
            && 10--20\% & $0.239\pm0.004$ & $0.00177\pm0.00019$  & 0.023 \\
            && 20--30\% & $0.219\pm0.004$ & $0.00216\pm0.00016$  & 0.016 \\
            && 30--40\% & $0.214\pm0.002$ & $0.00220\pm0.00008$  & 0.005 \\
            && 40--60\% & $0.212\pm0.007$ & $0.00202\pm0.00028$  & 0.062 \\
            && 60--80\% & $0.194\pm0.004$ & $0.00223\pm0.00016$  & 0.018 \\
\hline \hline
\end{tabular}%
\end{center}
}} }
\renewcommand{\baselinestretch}{1.}

\begin{figure}
\hskip-1.0cm \begin{center}
\includegraphics[width=15.0cm]{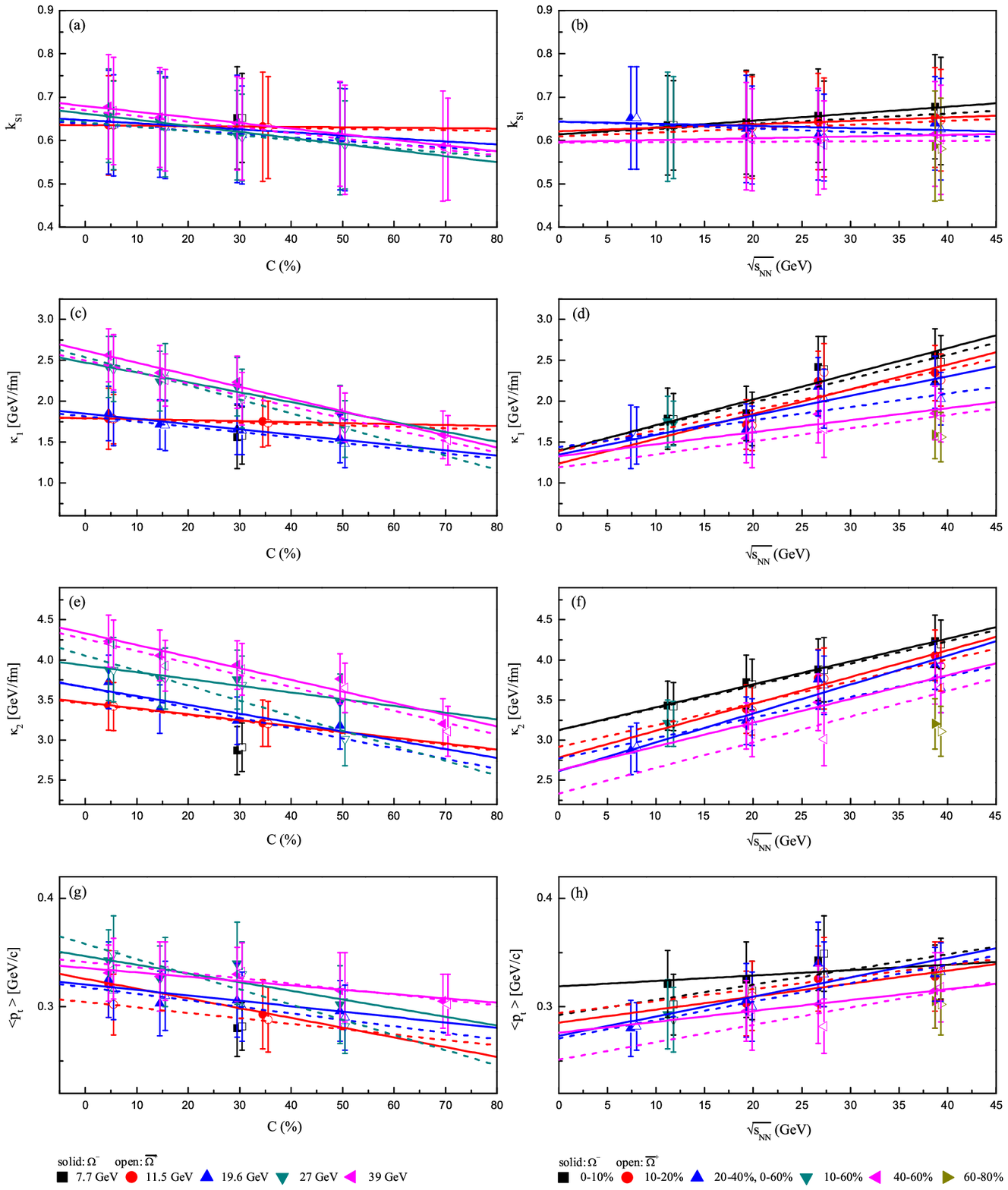}
\end{center}
\vskip.0cm Figure 5. As for Fig. 4, but showing the parameter
values extracted from Fig. 2 and listed in Table 2, where the
two-component Schwinger mechanism and single Erlang distribution
are used for $\Omega^-$ (solid symbols) and $\bar\Omega^+$ (open
symbols) spectra respectively. The solid and open symbols are
shifted respectively to the left and right sides by a small amount
for clearness.
\end{figure}

\renewcommand{\baselinestretch}{0.4}
\newpage
{\small {Table 5. Values of intercepts, slopes, and $\chi^2$/dof
corresponding to the lines in Fig. 5 which shows dependences of
parameters on centrality at different energies (a,c,e,g) and on
energy in different centrality intervals (b,d,f,h). The values of
parameters are extracted from Fig. 2 and listed in Table 2.
{%
\begin{center}
\begin{tabular}{cccccc}
\hline\hline  Figure & Relation & Type & Intercept & Slope & $\chi^2$/dof \\
\hline
& & & \multicolumn{3}{c}{Two-component Schwinger mechanism in Fig. 2} \\
\cline{4-6}
Figure 5(a) & $k_{S1}-C$ & 11.5 GeV $\Omega^{-}$       & $0.636\pm0.000$ & $-0.00009\pm0.00000$ & 0.001 \\
     && 11.5 GeV $\bar{\Omega}^{+}$ & $0.636\pm0.000$ & $-0.00017\pm0.00000$ & 0.001 \\
     && 19.6 GeV $\Omega^{-}$       & $0.647\pm0.001$ & $-0.00072\pm0.00004$ & 0.001 \\
     && 19.6 GeV $\bar{\Omega}^{+}$ & $0.639\pm0.002$ & $-0.00077\pm0.00007$ & 0.001 \\
     && 27 GeV   $\Omega^{-}$       & $0.662\pm0.005$ & $-0.00136\pm0.00017$ & 0.005 \\
     && 27 GeV   $\bar{\Omega}^{+}$ & $0.642\pm0.003$ & $-0.00106\pm0.00009$ & 0.001 \\
     && 39 GeV   $\Omega^{-}$       & $0.680\pm0.003$ & $-0.00132\pm0.00008$ & 0.002 \\
     && 39 GeV   $\bar{\Omega}^{+}$ & $0.670\pm0.003$ & $-0.00132\pm0.00007$ & 0.001 \\
Figure 5(b) & $k_{S1}-\sqrt{s_{NN}}$ & 0--10\%  $\Omega^{-}$       & $0.614\pm0.003$ & $0.00161\pm0.00013$  & 0.001 \\
     && 0--10\%  $\bar{\Omega}^{+}$ & $0.614\pm0.012$ & $0.00119\pm0.00044$  & 0.011 \\
     && 10--20\% $\Omega^{-}$       & $0.621\pm0.001$ & $0.00082\pm0.00004$  & 0.001 \\
     && 10--20\% $\bar{\Omega}^{+}$ & $0.609\pm0.008$ & $0.00093\pm0.00027$  & 0.002 \\
     && 20--40\% $\Omega^{-}$       & $0.644\pm0.018$ & $-0.00047\pm0.00071$ & 0.033 \\
     && 20--40\% $\bar{\Omega}^{+}$ & $0.644\pm0.019$ & $-0.00083\pm0.00074$ & 0.034 \\
     && 40--60\% $\Omega^{-}$       & $0.597\pm0.017$ & $0.00037\pm0.00057$  & 0.008 \\
     && 40--60\% $\bar{\Omega}^{+}$ & $0.595\pm0.015$ & $0.00010\pm0.00050$  & 0.008 \\
Figure 5(c) & $\kappa_1-C$ & 11.5 GeV $\Omega^{-}$       & $1.794\pm0.000$ & $-0.00117\pm0.00000$ & 0.001 \\
     && 11.5 GeV $\bar{\Omega}^{+}$ & $1.790\pm0.000$ & $-0.00167\pm0.00000$ & 0.001 \\
     && 19.6 GeV $\Omega^{-}$       & $1.851\pm0.027$ & $-0.00637\pm0.00090$ & 0.014 \\
     && 19.6 GeV $\bar{\Omega}^{+}$ & $1.818\pm0.014$ & $-0.00653\pm0.00046$ & 0.004 \\
     && 27 GeV   $\Omega^{-}$       & $2.474\pm0.043$ & $-0.01209\pm0.00144$ & 0.028 \\
     && 27 GeV   $\bar{\Omega}^{+}$ & $2.540\pm0.050$ & $-0.01718\pm0.00164$ & 0.034 \\
     && 39 GeV   $\Omega^{-}$       & $2.620\pm0.032$ & $-0.01480\pm0.00078$ & 0.021 \\
     && 39 GeV   $\bar{\Omega}^{+}$ & $2.500\pm0.029$ & $-0.01414\pm0.00071$ & 0.017 \\
Figure 5(d) & $\kappa_1-\sqrt{s_{NN}}$ & 0--10\%  $\Omega^{-}$       & $1.394\pm0.185$ & $0.03140\pm0.00702$  & 0.245 \\
     && 0--10\%  $\bar{\Omega}^{+}$ & $1.404\pm0.216$ & $0.02907\pm0.00823$  & 0.326 \\
     && 10--20\% $\Omega^{-}$       & $1.240\pm0.355$ & $0.03020\pm0.01200$  & 0.469 \\
     && 10--20\% $\bar{\Omega}^{+}$ & $1.393\pm0.537$ & $0.02505\pm0.01812$  & 1.115 \\
     && 20--40\% $\Omega^{-}$       & $1.350\pm0.164$ & $0.02392\pm0.00631$  & 0.276 \\
     && 20--40\% $\bar{\Omega}^{+}$ & $1.439\pm0.129$ & $0.01638\pm0.00497$  & 0.174 \\
     && 40--60\% $\Omega^{-}$       & $1.328\pm0.221$ & $0.01470\pm0.00746$  & 0.194 \\
     && 40--60\% $\bar{\Omega}^{+}$ & $1.191\pm0.071$ & $0.01603\pm0.00240$  & 0.021 \\
Figure 5(e) & $\kappa_2-C$ & 11.5 GeV $\Omega^{-}$       & $3.468\pm0.000$ & $-0.00730\pm0.00000$ & 0.001 \\
     && 11.5 GeV $\bar{\Omega}^{+}$ & $3.456\pm0.000$ & $-0.00727\pm0.00000$ & 0.001 \\
     && 19.6 GeV $\Omega^{-}$       & $3.659\pm0.027$ & $-0.01105\pm0.00333$ & 0.200 \\
     && 19.6 GeV $\bar{\Omega}^{+}$ & $3.658\pm0.014$ & $-0.01266\pm0.00266$ & 0.130 \\
     && 27 GeV   $\Omega^{-}$       & $3.930\pm0.043$ & $-0.00843\pm0.00160$ & 0.033 \\
     && 27 GeV   $\bar{\Omega}^{+}$ & $4.054\pm0.012$ & $-0.01874\pm0.00401$ & 0.214 \\
     && 39 GeV   $\Omega^{-}$       & $4.331\pm0.082$ & $-0.01449\pm0.00198$ & 0.147 \\
     && 39 GeV   $\bar{\Omega}^{+}$ & $4.260\pm0.085$ & $-0.01494\pm0.00204$ & 0.158 \\
Figure 5(f) & $\kappa_2-\sqrt{s_{NN}}$ & 0--10\%  $\Omega^{-}$       & $3.125\pm0.037$ & $0.02851\pm0.00140$  & 0.001 \\
     && 0--10\%  $\bar{\Omega}^{+}$ & $3.123\pm0.028$ & $0.02770\pm0.00105$  & 0.007 \\
     && 10--20\% $\Omega^{-}$       & $2.784\pm0.159$ & $0.03345\pm0.00536$  & 0.096 \\
     && 10--20\% $\bar{\Omega}^{+}$ & $2.915\pm0.228$ & $0.02717\pm0.00768$  & 0.184 \\
     && 20--40\% $\Omega^{-}$       & $2.613\pm0.135$ & $0.03600\pm0.00521$  & 0.179 \\
     && 20--40\% $\bar{\Omega}^{+}$ & $2.770\pm0.173$ & $0.02560\pm0.00667$  & 0.292 \\
     && 40--60\% $\Omega^{-}$       & $2.626\pm0.086$ & $0.02964\pm0.00291$  & 0.030 \\
     && 40--60\% $\bar{\Omega}^{+}$ & $2.336\pm0.349$ & $0.03187\pm0.01177$  & 0.537 \\
\hline
& & & \multicolumn{3}{c}{Single Erlang distribution in Fig. 2} \\
\cline{4-6}
Figure 5(g) & $\langle p_t \rangle-C$ & 11.5 GeV $\Omega^{-}$       & $0.326\pm0.000$ & $-0.00093\pm0.00000$ & 0.001 \\
     && 11.5 GeV $\bar{\Omega}^{+}$ & $0.304\pm0.000$ & $-0.00047\pm0.00000$ & 0.001 \\
     && 19.6 GeV $\Omega^{-}$       & $0.321\pm0.006$ & $-0.00052\pm0.00021$ & 0.075 \\
     && 19.6 GeV $\bar{\Omega}^{+}$ & $0.318\pm0.001$ & $-0.00056\pm0.00002$ & 0.001 \\
     && 27 GeV   $\Omega^{-}$       & $0.346\pm0.010$ & $-0.00076\pm0.00034$ & 0.155 \\
     && 27 GeV   $\bar{\Omega}^{+}$ & $0.358\pm0.008$ & $-0.00139\pm0.00027$ & 0.153 \\
     && 39 GeV   $\Omega^{-}$       & $0.336\pm0.003$ & $-0.00041\pm0.00007$ & 0.030 \\
     && 39 GeV   $\bar{\Omega}^{+}$ & $0.341\pm0.003$ & $-0.00051\pm0.00008$ & 0.030 \\
Figure 5(h) & $\langle p_t \rangle-\sqrt{s_{NN}}$ & 0--10\%  $\Omega^{-}$       & $0.319\pm0.010$ & $0.00046\pm0.00040$  & 0.121 \\
     && 0--10\%  $\bar{\Omega}^{+}$ & $0.292\pm0.018$ & $0.00135\pm0.00067$  & 0.264 \\
     && 10--20\% $\Omega^{-}$       & $0.285\pm0.016$ & $0.00118\pm0.00056$  & 0.130 \\
     && 10--20\% $\bar{\Omega}^{+}$ & $0.294\pm0.016$ & $0.00108\pm0.00055$  & 0.116 \\
     && 20--40\% $\Omega^{-}$       & $0.273\pm0.016$ & $0.00175\pm0.00060$  & 0.271 \\
     && 20--40\% $\bar{\Omega}^{+}$ & $0.271\pm0.010$ & $0.00173\pm0.00038$  & 0.129 \\
     && 40--60\% $\Omega^{-}$       & $0.276\pm0.002$ & $0.00100\pm0.00005$  & 0.001 \\
     && 40--60\% $\bar{\Omega}^{+}$ & $0.251\pm0.023$ & $0.00159\pm0.00077$  & 0.324 \\
\hline\hline
\end{tabular}%
\end{center}
}} }
\renewcommand{\baselinestretch}{1.}

\begin{figure}
\hskip-1.0cm \begin{center}
\includegraphics[width=15.0cm]{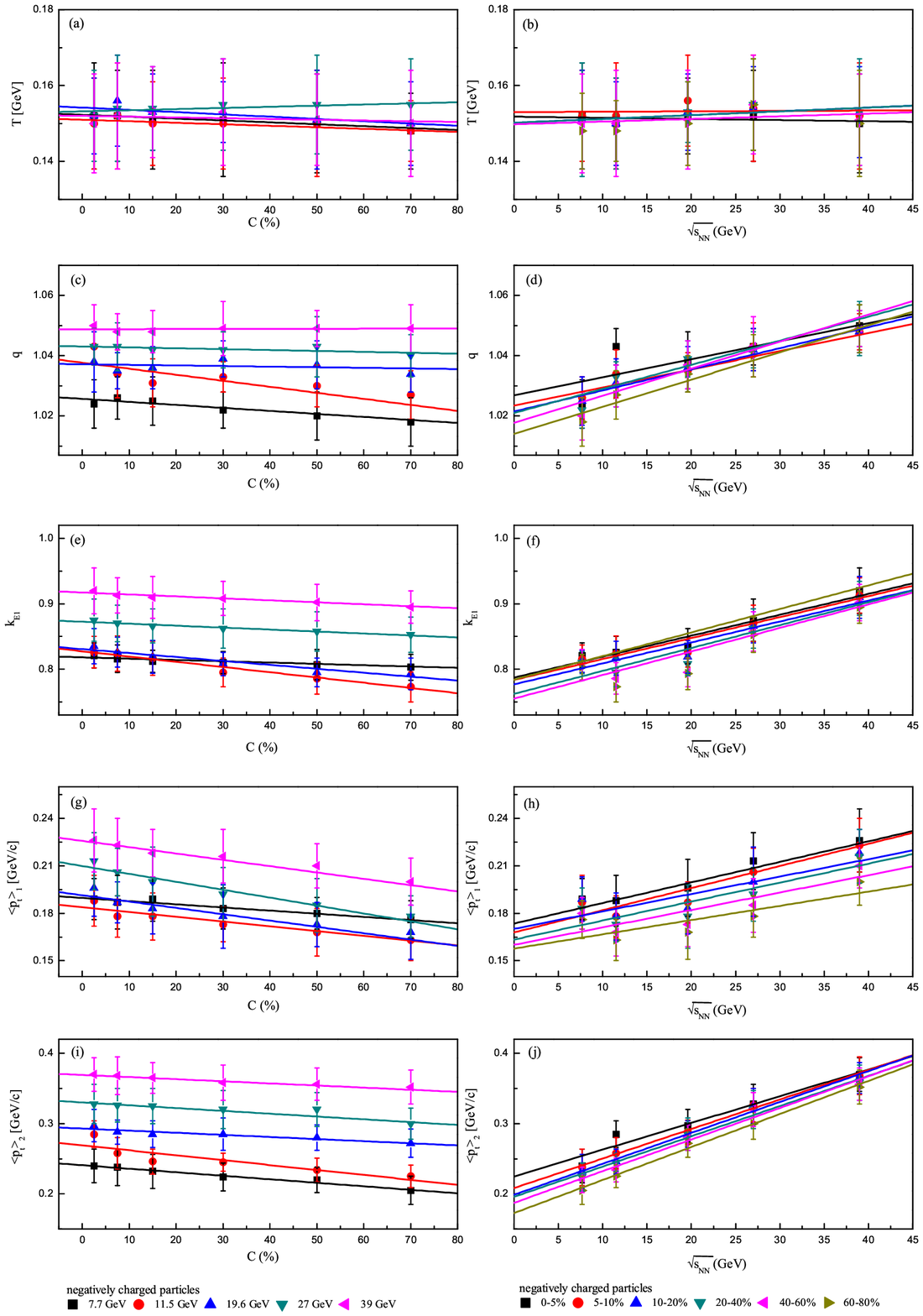}
\end{center}
\vskip.0cm Figure 6. Dependences of parameters (a,b) $T$, (c,d)
$q$, (e,f) $k_{E1}$, (g,h) $\langle p_t \rangle_1$, and (i,j)
$\langle p_t \rangle_2$ on centrality at different energies (left
panel) and on energy in different centrality intervals (right
panel). The different symbols represent the parameter values
extracted from Fig. 3 and listed in Table 3, where the single
Tsallis statistics and two-component Erlang distribution are used
for negatively charged particle spectra. The lines are our fitted
results.
\end{figure}

\renewcommand{\baselinestretch}{0.4}
\newpage
{\small {Table 6. Values of intercepts, slopes, and $\chi^2$/dof
corresponding to the lines in Fig. 6 which shows dependences of
parameters on centrality at different energies (a,c,e,g,i) and on
energy in different centrality intervals (b,d,f,h,j). The values
of parameters are extracted from Fig. 3 and listed in Table 3.
{%
\begin{center}
\begin{tabular}{cccccc}
\hline\hline  Figure & Relation & Type & Intercept & Slope & $\chi^2$/dof \\
\hline \hline
& & & \multicolumn{3}{c}{Single Tsallis statistics in Fig. 3} \\
\cline{4-6}
Figure 6(a) & $T-C$ & 7.7 GeV  & $0.152\pm0.001$ & $-0.00005\pm0.00001$ & 0.001 \\
     && 11.5 GeV & $0.151\pm0.001$ & $-0.00004\pm0.00001$ & 0.006 \\
     && 19.6 GeV & $0.154\pm0.001$ & $-0.00006\pm0.00001$ & 0.021 \\
     && 27 GeV   & $0.153\pm0.001$ & $0.00003\pm0.00001$  & 0.005 \\
     && 39 GeV   & $0.152\pm0.001$ & $-0.00002\pm0.00001$ & 0.012 \\
Figure 6(b) & $T-\sqrt{s_{NN}}$ & 0--5\%   & $0.152\pm0.001$ & $-0.00003\pm0.00004$ & 0.010 \\
     && 5--10\%  & $0.153\pm0.002$ & $0.00001\pm0.00007$  & 0.027 \\
     && 10--20\% & $0.150\pm0.001$ & $0.00010\pm0.00004$  & 0.014 \\
     && 20--40\% & $0.150\pm0.001$ & $0.00010\pm0.00006$  & 0.019 \\
     && 40--60\% & $0.150\pm0.002$ & $0.00007\pm0.00007$  & 0.028 \\
     && 60--80\% & $0.148\pm0.002$ & $0.00012\pm0.00010$  & 0.055 \\
Figure 6(c) & $q-C$ & 7.7 GeV  & $1.026\pm0.001$ & $-0.00011\pm0.00002$ & 0.019 \\
     && 11.5 GeV & $1.038\pm0.001$ & $-0.00017\pm0.00006$ & 0.322 \\
     && 19.6 GeV & $1.037\pm0.001$ & $-0.00002\pm0.00003$ & 0.098 \\
     && 27 GeV   & $1.043\pm0.001$ & $-0.00003\pm0.00001$ & 0.012 \\
     && 39 GeV   & $1.049\pm0.001$ & $0.00001\pm0.00001$  & 0.015 \\
Figure 6(d) & $q-\sqrt{s_{NN}}$ & 0--5\%   & $1.027\pm0.006$ & $0.00061\pm0.00024$  & 1.079 \\
     && 5--10\%  & $1.024\pm0.002$ & $0.00065\pm0.00009$  & 0.133 \\
     && 10--20\% & $1.022\pm0.001$ & $0.00071\pm0.00006$  & 0.053 \\
     && 20--40\% & $1.021\pm0.003$ & $0.00076\pm0.00014$  & 0.458 \\
     && 40--60\% & $1.018\pm0.003$ & $0.00086\pm0.00012$  & 0.131 \\
     && 60--80\% & $1.014\pm0.002$ & $0.00093\pm0.00009$  & 0.120 \\
\hline
& & & \multicolumn{3}{c}{Two-component Erlang distribution in Fig. 3} \\
\cline{4-6}
Figure 6(e) & $k_{E1}-C$ & 7.7 GeV  & $0.818\pm0.001$ & $-0.00023\pm0.00004$ & 0.017 \\
     && 11.5 GeV & $0.827\pm0.003$ & $-0.00082\pm0.00007$ & 0.040 \\
     && 19.6 GeV & $0.831\pm0.003$ & $-0.00062\pm0.00008$ & 0.053 \\
     && 27 GeV   & $0.873\pm0.001$ & $-0.00029\pm0.00003$ & 0.005 \\
     && 39 GeV   & $0.917\pm0.001$ & $-0.00032\pm0.00004$ & 0.005 \\
Figure 6(f) & $k_{E1}-\sqrt{s_{NN}}$ & 0--5\%   & $0.787\pm0.009$ & $0.00325\pm0.00038$  & 0.128 \\
     && 5--10\%  & $0.784\pm0.011$ & $0.00316\pm0.00045$  & 0.315 \\
     && 10--20\% & $0.777\pm0.012$ & $0.00324\pm0.00050$  & 0.383 \\
     && 20--40\% & $0.762\pm0.016$ & $0.00354\pm0.00067$  & 1.138 \\
     && 40--60\% & $0.755\pm0.020$ & $0.00356\pm0.00082$  & 1.094 \\
     && 60--80\% & $0.784\pm0.020$ & $0.00359\pm0.00084$  & 1.218 \\
Figure 6(g) & $\langle p_t \rangle_1-C$ & 7.7 GeV  & $0.190\pm0.001$ & $-0.00020\pm0.00002$ & 0.009 \\
     && 11.5 GeV & $0.184\pm0.002$ & $-0.00031\pm0.00005$ & 0.049 \\
     && 19.6 GeV & $0.191\pm0.002$ & $-0.00036\pm0.00005$ & 0.044 \\
     && 27 GeV   & $0.210\pm0.002$ & $-0.00049\pm0.00005$ & 0.035 \\
     && 39 GeV   & $0.226\pm0.001$ & $-0.00035\pm0.00003$ & 0.014 \\
Figure 6(h) & $\langle p_t \rangle_1-\sqrt{s_{NN}}$ & 0--5\%   & $0.175\pm0.003$ & $0.00129\pm0.00014$  & 0.064 \\
     && 5--10\%  & $0.168\pm0.006$ & $0.00135\pm0.00026$  & 0.263 \\
     && 10--20\% & $0.170\pm0.007$ & $0.00112\pm0.00030$  & 0.349 \\
     && 20--40\% & $0.163\pm0.008$ & $0.00119\pm0.00032$  & 0.383 \\
     && 40--60\% & $0.160\pm0.008$ & $0.00110\pm0.00034$  & 0.706 \\
     && 60--80\% & $0.158\pm0.008$ & $0.00092\pm0.00033$  & 0.517 \\
Figure 6(i) & $\langle p_t \rangle_2-C$ & 7.7 GeV  & $0.241\pm0.001$ & $-0.00049\pm0.00004$ & 0.016 \\
     && 11.5 GeV & $0.369\pm0.007$ & $-0.00069\pm0.00018$ & 0.393 \\
     && 19.6 GeV & $0.293\pm0.002$ & $-0.00029\pm0.00005$ & 0.029 \\
     && 27 GeV   & $0.330\pm0.003$ & $-0.00035\pm0.00007$ & 0.041 \\
     && 39 GeV   & $0.369\pm0.001$ & $-0.00026\pm0.00003$ & 0.006 \\
Figure 6(j) & $\langle p_t \rangle_2-\sqrt{s_{NN}}$ & 0--5\%   & $0.225\pm0.010$ & $0.00377\pm0.00044$  & 0.376 \\
     && 5--10\%  & $0.208\pm0.003$ & $0.00416\pm0.00014$  & 0.029 \\
     && 10--20\% & $0.199\pm0.005$ & $0.00438\pm0.00019$  & 0.059 \\
     && 20--40\% & $0.196\pm0.006$ & $0.00432\pm0.00024$  & 0.082 \\
     && 40--60\% & $0.187\pm0.007$ & $0.00452\pm0.00030$  & 0.156 \\
     && 60--80\% & $0.173\pm0.004$ & $0.00468\pm0.00018$  & 0.069 \\
\hline\hline
\end{tabular}%
\end{center}
}} }
\renewcommand{\baselinestretch}{1.}

\begin{figure}
\hskip-1.0cm \begin{center}
\includegraphics[width=15.0cm]{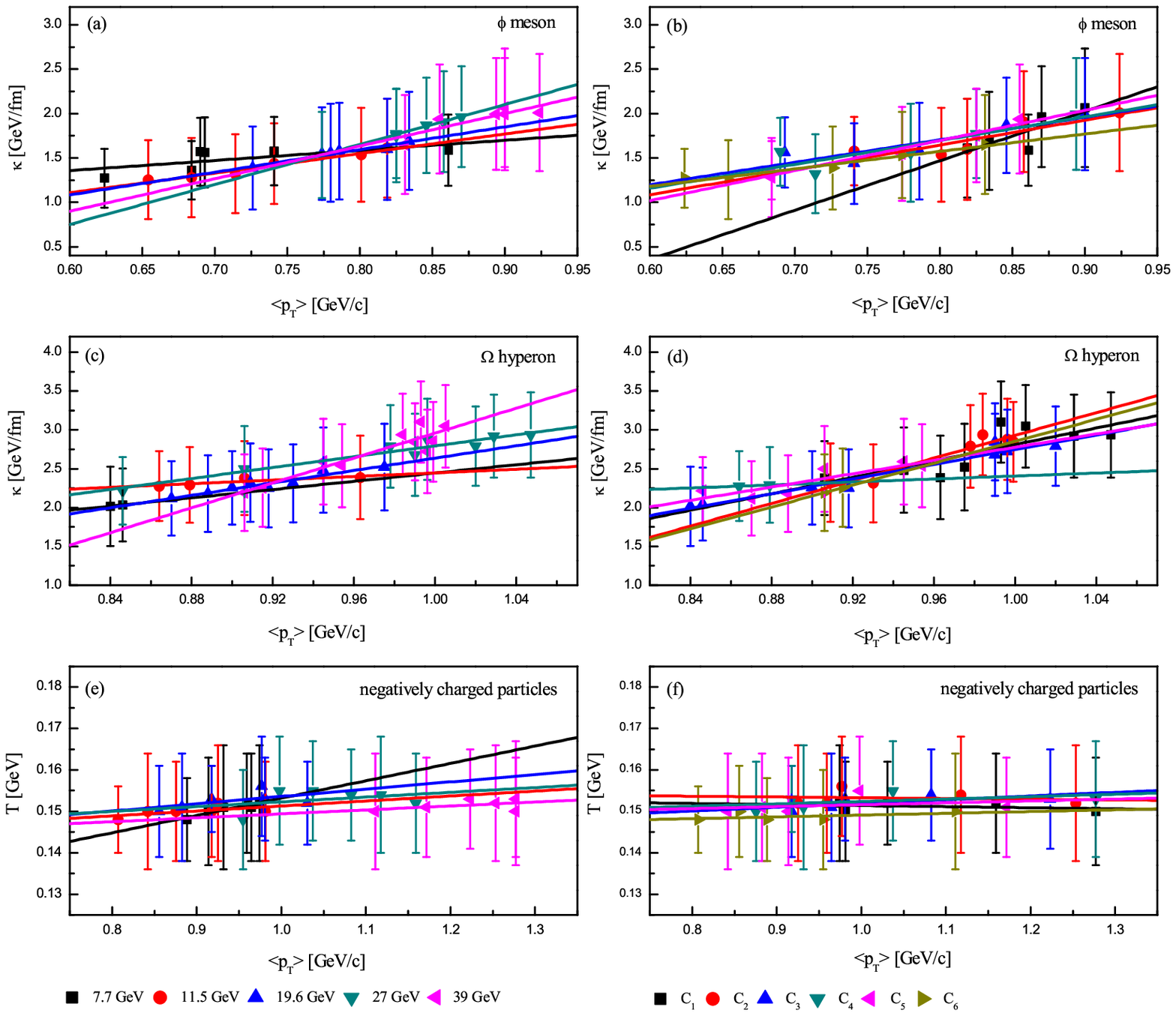}
\end{center}
\vskip.0cm Figure 7. Relations between different parameters, (a,b)
$\kappa$ and $\langle p_T \rangle$ for $\phi$ mesons, (c,d)
$\kappa$ and $\langle p_T \rangle$ for $\Omega$ hyperons, and
(e,f) $T$ and $\langle p_T \rangle$ for negatively charged
particles. The different symbols represent the parameter values at
different energies (in various centrality intervals) (left panel)
and in different centrality intervals (at various energies) (right
panel), where $C_1$, $C_2$, $\cdots$, $C_6$ for Figures 7(b),
7(d), and 7(f) represent the centrality intervals mentioned in the
right panels of Figures 4, 5, and 6, respectively. In some cases
the parameter values are obtained by an average weighted by two
components which are extracted from Figs. 1--3 and listed in
Tables 1--3. The lines are our fitted results.
\end{figure}

\renewcommand{\baselinestretch}{0.4}
\newpage
{\small {Table 7. Values of intercepts, slopes, and $\chi^2$/dof
corresponding to the lines in Fig. 7 which shows the relations
between different parameters.
{%
\begin{center}
\begin{tabular}{cccccc}
\hline\hline  Figure & Relation & Type & Intercept & Slope & $\chi^2$/dof \\
\hline
Figure 7(a) & $\kappa-\langle p_T \rangle$ & 7.7 GeV  & $0.674\pm0.406$  & $1.139\pm0.564$ & 0.098 \\
     && 11.5 GeV & $-0.218\pm0.133$ & $2.208\pm0.180$  & 0.004 \\
     && 19.6 GeV & $-0.448\pm0.239$ & $2.553\pm0.304$  & 0.003 \\
     && 27 GeV   & $-1.951\pm0.092$ & $4.502\pm0.110$  & 0.001 \\
     && 39 GeV   & $-1.305\pm0.906$ & $3.672\pm0.024$  & 0.021 \\
Figure 7(b) & $\kappa-\langle p_T \rangle$ & 0--10\%  & $-2.984\pm1.712$ & $5.565\pm1.997$ & 0.117 \\
     && 10--20\% & $-0.596\pm0.628$ & $2.802\pm0.775$  & 0.057 \\
     && 20--30\% & $-0.323\pm0.502$ & $2.534\pm0.630$  & 0.072 \\
     && 30--40\% & $-0.452\pm0.575$ & $2.864\pm0.733$  & 0.111 \\
     && 40--60\% & $-1.016\pm0.262$ & $3.391\pm0.341$  & 0.020 \\
     && 60--80\% & $-0.001\pm0.153$ & $1.966\pm0.211$  & 0.011 \\
Figure 7(c) & $\kappa-\langle p_T \rangle$ & 7.7 GeV  & $-0.221\pm0.001$ & $2.666\pm0.001$ & 0.001 \\
     && 11.5 GeV & $1.274\pm0.342$  & $1.174\pm0.379$  & 0.005 \\
     && 19.6 GeV & $-1.353\pm0.446$ & $3.990\pm0.486$  & 0.008 \\
     && 27 GeV   & $-0.715\pm0.361$ & $3.511\pm0.369$  & 0.019 \\
     && 39 GeV   & $-5.062\pm1.010$ & $8.021\pm1.041$  & 0.052 \\
Figure 7(d) &$\kappa-\langle p_T \rangle$& 0--10\%    & $-2.481\pm1.539$ & $5.294\pm1.564$ & 0.148 \\
     && 10--20\% & $-4.378\pm0.991$ & $7.308\pm1.025$  & 0.035 \\
     && 20--40\% & $-1.980\pm0.370$ & $4.726\pm0.393$  & 0.026 \\
     && 10--60\% & $1.447\pm0.001$  & $0.959\pm0.001$  & 0.001 \\
     && 40--60\% & $-1.496\pm0.986$ & $4.273\pm1.093$  & 0.054 \\
     && 60--80\% & $-4.203\pm0.022$ & $7.058\pm0.025$  & 0.001 \\
Figure 7(e) & $T-\langle p_T \rangle$ & 7.7 GeV  & $0.111\pm0.007$  & $0.042\pm0.008$ & 0.003 \\
     && 11.5 GeV & $0.139\pm0.006$  & $0.012\pm0.007$  & 0.010 \\
     && 19.6 GeV & $0.136\pm0.011$  & $0.018\pm0.012$  & 0.031 \\
     && 27 GeV   & $0.141\pm0.016$  & $0.012\pm0.015$  & 0.054 \\
     && 39 GeV   & $0.140\pm0.010$  & $0.010\pm0.008$  & 0.011 \\
Figure 7(f) & $T-\langle p_T \rangle$ & 0--5\%   & $0.154\pm0.004$  & $-0.003\pm0.004$ & 0.010 \\
     && 5--10\%  & $0.155\pm0.007$  & $-0.002\pm0.006$ & 0.027 \\
     && 10--20\% & $0.143\pm0.005$  & $0.009\pm0.005$  & 0.017 \\
     && 20--40\% & $0.147\pm0.005$  & $0.006\pm0.005$  & 0.029 \\
     && 40--60\% & $0.147\pm0.007$  & $0.005\pm0.008$  & 0.032 \\
     && 60--80\% & $0.145\pm0.004$  & $0.004\pm0.004$  & 0.010 \\
\hline \hline
\end{tabular}%
\end{center}
}} }


\newpage

\begin{thebibliography}{99}

\bibitem{1}
S. Chatterjee, S. Das, L. Kumar, D. Mishra, B. Mohanty, R. Sahoo,
N. Sharma, Adv. High Energy Phys. {\bf 2015}, 349013 (2015).
\bibitem{2}
Y. Zhong, C.-B. Yang, X. Cai, S.-Q. Feng, Adv. High Energy Phys.
{\bf 2015}, 193039 (2015).
\bibitem{3}
J. Uphoff, O. Fochler, Z. Xu, C. Greiner, Acta Phys. Pol. B Proc.
Suppl. {\bf 5}, 555 (2012).
\bibitem{4}
R. C. Hwa, Adv. High Energy Phys. {\bf 2015}, 526908 (2015).
\bibitem{5}
G.-L. Ma, M.-W. Nie, Adv. High Energy Phys. {\bf 2015}, 967474
(2015).
\bibitem{6}
D. D. Ivanenko, D. F. Kurdgelaidze, Astrofizika {\bf 1}, 479
(1965) [Astrophysics {\bf 1}, 251 (1965)].
\bibitem{7}
N. Itoh, Prog. Theor. Phys. {\bf 44} 291, (1970).
\bibitem{8}
T. D. Lee, G. C. Wick, Phys. Rev. D {\bf 9}, 2291 (1974).
\bibitem{9}
M. Nasim, V. Bairathi, M. K. Sharma, B. Mohanty, A. Bhasin, Adv.
High Energy Phys. {\bf 2015}, 197930 (2015).
\bibitem{10}
For the STAR Collaboration (L. Kumar), {\it Talk given the 2011
Meeting of the Division of Particles and Fields of the American
Physical Society (DPF 2011), Rhode Island, USA, August 9-13,
2011}, arXiv:1109.5313 (2011).
\bibitem{11}
For the STAR collaboration (S. Das), EPJ Web of Conf. {\bf 90},
08007 (2015).
\bibitem{12}
For the STAR collaboration (S. S. Shi), Acta Phys. Pol. B Proc.
Suppl. {\bf 5}, 311 (2012).
\bibitem{13}
For the NA49 NA61/SHINE Collaborations (M. Gazdzicki), J. Phys. G
{\bf 38}, 124024 (2011).
\bibitem{14}
For the NA61/SHINE Collaboration (M. Mackowiak-Pawlowska), PoS
(EPS-HEP2015), 205 (2015), arXiv:1510.08688 (2015).
\bibitem{15}
For the NA61 Collaboration (M. Mackowiak), J. Phys. Conf. Ser.
{\bf 270}, 012048 (2011).
\bibitem{16}
For the NA61/SHINE Collaboration (D. T. Larsen), arXiv:1510.00674
(2015).
\bibitem{17}
For the STAR Collaboration (L. Kumar), J. Phys. G {\bf 38}, 124145
(2011).
\bibitem{18}
A. Rustamov, Cen. Eur. J. Phys. {\bf 10}, 1267 (2012).
\bibitem{19}
M. Bleicher, arXiv:hep-ph/0509314 (2005).
\bibitem{20}
J. Steinheimer, M. Bleicher, Eur. Phys. J. A {\bf 48}, 100 (2012).
\bibitem{21}
Y. B. Ivanov, D. Blaschke, Phys. Rev. C {\bf 92} 024916 (2015).
\bibitem{22}
Y. B. Ivanov, {\it Talk given at the 15th International Conference
on Strangeness in Quark Matter (SQM2015), Dubna, Russia, 6-11 July
2015}, arXiv:1509.06944 (2015).
\bibitem{23}
F.-H. Liu, L.-N. Gao, R. A. Lacey, Adv. High Energy Phys. {\bf
2016}, 9467194 (2016).
\bibitem{24}
R. A. Lacey, Phys. Rev. Lett. {\bf 114}, 142301 (2015).
\bibitem{25}
F.-H. Liu, Y.-Q. Gao, T. Tian, B.-C. Li, Eur. Phys. J. A {\bf 50},
94 (2014).
\bibitem{26}
F.-H. Liu, J.-S. Li, Phys. Rev. C {\bf 78}, 044602 (2008).
\bibitem{27}
F.-H. Liu, Nucl. Phys. A {\bf 810}, 159 (2008).
\bibitem{28}
J. Schwinger, Phys. Rev. {\bf 82}, 664 (1951).
\bibitem{29}
R.-C. Wang, C.-Y. Wong, Phys. Rev. D {\bf 38}, 348 (1988).
\bibitem{30}
C.-Y. Wong, {\it Introduction to High Energy Heavy Ion
Collisions}, World Scientific, Singapore, 1994.
\bibitem{31}
P. Braun-Munzinger, K. Redlich, J. Stachel, in {\it Quark-Gluon
Plasma 3}, edited by R. C. Hwa, X.-N. Wang (World Scientific,
Singapore, 2004), arXiv:nucl-th/0304013 (2004).
\bibitem{32}
C. Tsallis, J. Stat. Phys. {\bf 52}, 479 (1988).
\bibitem{33}
T. S. Bir{\'o}, G. Purcsel, K. {\"U}rm{\"o}ssy, Eur. Phys. J. A
{\bf 40}, 325 (2009).
\bibitem{34}
J. Cleymans, D. Worku, Eur. Phys. J. A {\bf 48}, 160 (2012).
\bibitem{35}
STAR Collaboration (L. Adamczyk {\it et al.}), arXiv:1506.07605
(2015).
\bibitem{36}
For the STAR Collaboration (M. V. Tokarev), {\it Talk given at the
International Conference ``HADRON STRUCTURE'15", Horny Smokovec,
Slovak Republic, 29 June -- 3 July, 2015}, arXiv:1509.06570,
(2015).
\bibitem{37}
A. Andronic, P. Braun-Munzinger, J. Stachel, Nucl. Phys. A {\bf
834}, 237c (2010).
\bibitem{38}
F.-H. Liu, Y.-Q. Gao, H.-R. Wei, Adv. High Energy Phys. {\bf
2014}, 293873 (2014).
\bibitem{39}
L.-N. Gao, F.-H. Liu, Adv. High Energy Phys. {\bf 2016}, 1505823
(2016).
\bibitem{40}
H.-R. Wei, Y.-H. Chen, L.-N. Gao, F.-H. Liu, Adv. High Energy
Phys. {\bf 2014}, 782631 (2014).
\end{thebibliography}
\end{document}